\documentclass[preprint,12pt]{elsarticle}
\usepackage[monochrome]{xcolor}
\biboptions{numbers,sort&compress}
\usepackage[utf8]{inputenc}
\usepackage[T1]{fontenc}
\usepackage{makecell}
\usepackage{nicematrix}
\usepackage{enumitem}
\usepackage{amsthm}
\usepackage{amssymb}
\usepackage{bm}
\usepackage{float}
\usepackage{mathrsfs}
\usepackage{amsfonts}
\usepackage{graphicx}
\usepackage{algorithm, algpseudocode}
\usepackage{subfigure}
\usepackage{tabularx}
\usepackage{flushend}
\usepackage{epsfig}
\usepackage[super]{nth}

\usepackage{soul}
\usepackage{epstopdf}
\usepackage{multirow}
\usepackage{hyperref}

\newcommand{\R}{\mathbb{R}}

\def\TR{\text{TransfoRhythm}}

\journal{Applied Soft Computing}
\begin{document}
\begin{frontmatter}

\title{$\TR$: A Transformer Architecture Conductive to Blood Pressure Estimation through Solo PPG Signal Capturing}

\author[1]{Amir Arjomand}
\author[2]{Amin Boudesh}
\author[2]{Farnoush Bayatmakou}
\author[1]{Georgiy Krylov}
\author[1,*]{Kenneth B. Kent}
\author[2,*]{Arash Mohammadi}
\affiliation[1]{organization={Faculty of Computer Science}, addressline={University of New Brunswick}, city={Fredericton}, state={New Brunswick}, country={Canada}}
\affiliation[2]{organization={Concordia Institute for Information Systems Engineering}, addressline={Concordia University}, city={Montreal}, state={Quebec}, country={Canada}}

\begin{abstract} Recent statistics indicate that approximately $1.3$ billion individuals worldwide are affected by hypertension, a major contributor to premature mortality. Blood pressure (BP) is an important physiological indicator for hypertension assessment and management. Conventional cuff-based methods provide intermittent measurements and may be unsuitable for continuous monitoring. Recent advances in artificial intelligence (AI) and deep neural networks (DNNs) have therefore increased interest in data-driven cuff-less BP estimation. Existing studies frequently combine electrocardiography (ECG) and photoplethysmography (PPG), requiring multiple synchronized sensor types. PPG-only estimation offers a simpler acquisition setup but remains challenging because BP-related morphological information must be extracted in the presence of motion artifacts and high-frequency noise. This paper introduces TransfoRhythm, a Transformer-based DNN framework for PPG-only BP estimation using the MIMIC-IV Waveform Database. The framework uses multi-head attention to model temporal relationships among cycle-level PPG features. Evaluation was performed using patient-independent 5-fold cross-validation, with patient IDs used as the splitting unit and zero patient overlap between training and test folds. Within each fold, feature selection, collinearity filtering, data scaling, hyperparameter selection, and model fitting were performed using only the training patients. \textcolor{orange}{ Under this protocol, TransfoRhythm achieved mean MAEs of $1.93$~mmHg for systolic BP and $1.49$~mmHg for diastolic BP, with corresponding RMSE values of $3.11$ and $2.59$~mmHg. The mean errors were $-0.01$~mmHg and $-0.26$~mmHg, while the mean $R^2$ values were $0.888$ and $0.891$ for systolic and diastolic BP, respectively. These findings demonstrate consistent predictive performance across held-out patient groups within the evaluated MIMIC-IV cohort. External and prospective validation remains necessary to establish generalizability beyond this dataset. } 
\end{abstract}



\end{frontmatter}


\section{Introduction} \label{sec:intro}
High blood pressure is a serious symptom of hypertensive disease that can lead to disability and even death \citep{1-zabihi2022bp}. Based on an estimate by the American Heart Association, approximately $41.4$ \% of US people will be struggling with hypertension by $2030$~\citep{2-heidenreich2011forecasting}. There are various approaches to measuring Blood Pressure, among which, cuff-based measurement~\citep{3-lewis2019oscillometric} (where a cuff is employed to temporarily stop the blood flow to obtain the extremum BP values) is the common traditional method used both in hospitals and in-home care. Such an approach, however, carries the following drawbacks: (i) Cuff devices are not readily accessible in resource-limited settings, leading to the lack of awareness and control of hypertensive conditions; (ii) The inflation and deflation of cuffs can disrupt gold-standard hypertension diagnosis and hypotension surveillance, making these devices less suitable for frequent or prolonged usage, and; (iii) Widely used cuff-based devices fail to provide continuous BP measurement, limiting immediate detection of hypotension and real-time therapy titration~\citep{4-mukkamala2022cuffless}.  Alternative methods of BP measurement, such as cuff-less monitoring, have been explored to provide continuous monitoring and reduce the potential for measurement errors caused by cuff-based methods. Cuffless BP measurement offers the potential for continuous monitoring, which enables long-term hypertension management and facilitates the precise diagnosis of cardiovascular conditions.~\citep{4-mukkamala2022cuffless}. Continuous BP monitoring offers significant advantages, particularly in detecting nocturnal BP patterns. Nocturnal BP provides critical information for the risk assessment of cardiovascular diseases (CVD). For instance, abnormally high surges or deep dips in BP during the night (more than $10-20$\%) are strongly linked to severe health outcomes, including an increased possibility of stroke, heart failure, and even mortality~\citep{5-kario2022effect}. These fluctuations may go unnoticed with traditional cuff-based measurements, which lack the ability to continuously track BP over time.
The history of cuff-less BP estimation arises from arterial stiffness measurement through Pulse Wave Velocity (PWV)~\citep{7-shao2020optimization,8-yousefian2020pulse,9-esmaili2017nonlinear,10-pielmus2021surrogate}. In $1981$, Geddes and Voelz tried, for the first time, to analyze Pulse Transit Time (PTT) and BP on ten dogs and reported a meaningful correlation between PTT and diastolic BP~\citep{11-geddes1981pulse}. Recent cuffless measurement techniques typically employ biological signals such as Photoplethysmography (PPG), Ballistocardiography (BCG), and Electrocardiography (ECG) to estimate BP. PPG is an optical technique used to measure volumetric variations in blood circulation, offering valuable insights into the cardiovascular system. The method relies on emitting and detecting near-infrared light (wavelengths of 800–2500 nm) reflected from arterial blood flow. Changes in light absorption, caused by pulsatile blood flow through vessels, are captured by a sensor placed on the skin and can be employed for BP estimation. BCG detects body movements caused by cardiac contractions using a sensor, typically placed on a seat, which can be used for BP estimation. ECG measures the heart's electrical activities, which can be used to derive various parameters related to cardiac function, including BP~\citep{12-mishra2017cuffless}. 
\textcolor{orange}{ Recent advances in machine learning (ML), particularly deep learning (DL), have increased interest in cuff-less BP estimation from ECG and PPG signals. Existing multimodal approaches employ different architectures to capture complementary morphological and temporal information. For example, Baker \textit{et al.}~\citep{13-baker2021hybrid} combined CNN-based feature extraction with LSTM-based temporal modeling, reporting MAEs of 4.41~mmHg for SBP and 2.91~mmHg for DBP on data from 84 subjects. Temporal and feed-forward alternatives have also been investigated: the TCN proposed in~\citep{1-zabihi2022bp} achieved RMSEs of 3.03~mmHg for SBP and 1.58~mmHg for DBP on the MIMIC-I and MIMIC-III datasets, while the modified MLP-Mixer introduced by Huang \textit{et al.}~\citep{14-huang2022mlp} achieved RMSEs of 5.10~mmHg and 3.13~mmHg on MIMIC-II. Collectively, these studies demonstrate that CNN--recurrent, temporal-convolutional, and MLP-based architectures can capture BP-related information from physiological waveforms. However, their dependence on both ECG and PPG requires multiple synchronized sensors, increasing acquisition complexity and motivating the development of reliable PPG-only approaches. } 

Although BP estimation via PPG and ECG sensors is promising, it suffers from some critical drawbacks, such as reliance on two types of sensors, high power consumption, the need for sensor re-calibration, and dependence on keeping a fixed distance between sensors. A promising solution to these issues lies in using stand-alone PPG signals. Due to the ease of use and potential for BP estimation, PPG sensors have become increasingly popular in the development of advanced signal processing/learning models ~\citep{18-le2020continuous,19-welykholowa2020multimodal,20-bassiouni2021combination,21-chao2021machine}. In this context, when it comes to feature extraction from PPG signals, two primary approaches are employed (as the appropriate feed for downstream processing/learning modules), i.e., raw PPG signals (time series input)~\citep{14-huang2022mlp,22-qin2021deep,13-baker2021hybrid,23-mahmud2023nabnet,tian2025paralleled,zhao2024amrunet,guo2024cuffless,zhao2024lightweightcnn,dai2024noninvasive}, and feature-based vectors~\citep{24-hassani2019improved,25-el2020cuffless,26-hsu2020generalized,27-el2021cuffless,28-gupta2022higher}. These representations can be extracted from the original PPG signal, multiple-order signal derivative, and Pulse Wave Decomposition (PWD) of the PPG. In the modeling step, to achieve accurate BP estimation, recent studies have employed various methodologies ranging from traditional ML models, such as Support Vector Regression (SVR) and Random Forest~\citep{28-gupta2022higher,29-liu2017cuffless,30-rastegar2023hybrid}, to Deep Neural Network (DNN) architectures~\citep{1-zabihi2022bp,31-paviglianiti2022comparison,zhao2024amrunet,dai2024noninvasive} such as LSTM networks, and Gated Recurrent Unit (GRU) models with attention mechanisms~\citep{23-mahmud2023nabnet,27-el2021cuffless,32-el2021deep,guo2024cuffless}. Lately, transformer-based networks have emerged as major players in this field and there are a handful of recent studies~\citep{tian2025paralleled,dai2024noninvasive,ma2022kd,huang2023arterialnet, ma2024stp, nawaz2024cufflessarterialbloodpressure} that investigated the potential of the attention mechanism for BP estimation through PPG signals.\textcolor{blue}{in this context, preliminary raw-PPG experiments produced lower validation performance and required additional feature-learning layers, increasing model complexity. Therefore, the current framework uses compact PPG, FDPPG, and SDPPG features, while end-to-end raw-waveform learning remains future work. }  \textcolor{orange}{ Recent PPG-only BP-estimation studies increasingly use attention-based architectures to model temporal dependencies without requiring an auxiliary ECG signal. These methods can be grouped into several complementary directions. Nawaz \textit{et al.}\citep{nawaz2024cufflessarterialbloodpressure} combined Transformer-based learning with frequency-domain linear/nonlinear regression on the UCL dataset\citep{UCLData}, with the frequency-domain model achieving better performance. Other studies focused on learning stronger waveform representations: Huang \textit{et al.}\citep{huang2023arterialnet} proposed ArterialNet, Ma \textit{et al.}\citep{ma2024stp} investigated self-supervised Transformer-based transfer learning, and Tian \textit{et al.}\citep{tian2025paralleled} combined CNN-based local feature extraction with Transformer-based temporal modeling on MIMIC-III. Collectively, these studies demonstrate the potential of attention, frequency-domain modeling, self-supervised learning, and hybrid architectures for PPG-only BP estimation. However, differences in datasets, input representations, and evaluation protocols limit direct comparison across studies. Moreover, removing the ECG signal places greater importance on preserving informative PPG morphology while controlling motion artifacts and high-frequency noise\citep{17-wang2021cuff}. These limitations motivate a patient-independent evaluation of morphology-aware, PPG-only BP estimation on MIMIC-IV. }
To address the challenges mentioned above, the paper introduces the $\TR$ framework, an attention-based DNN architecture built upon the recently released physiological database, MIMIC-IV dataset~\citep{33-johnson2020mimic}. To our knowledge,  the paper represents the first study to apply the MIMIC IV dataset for BP prediction. The $\TR$ framework is built on the regressive time series transformer network, which is crafted to learn from time series data while preserving sequence order and allowing parallel computation during training and inference. To expand the feature vectors' dimensionality, an embedding feature similar to those used in Natural Language Processing (NLP) tasks is employed. Positional encoding is then added to the extracted features to create the input data matrix, which is fed into a model block via the MHA mechanism. The output is refined in a position-wise block, which is then processed by a time compressor and flattening blocks to shorten the time frame and calculate estimated systolic and diastolic BP values. {\color{orange}
	The obtained results demonstrate the effectiveness of the proposed framework
	under the current retrospective evaluation protocol and motivate further
	external and prospective validation.
}

\section{Methodology} \label{sec:materials}
\begin{figure}[t!]
\centering
\includegraphics[width=\linewidth]{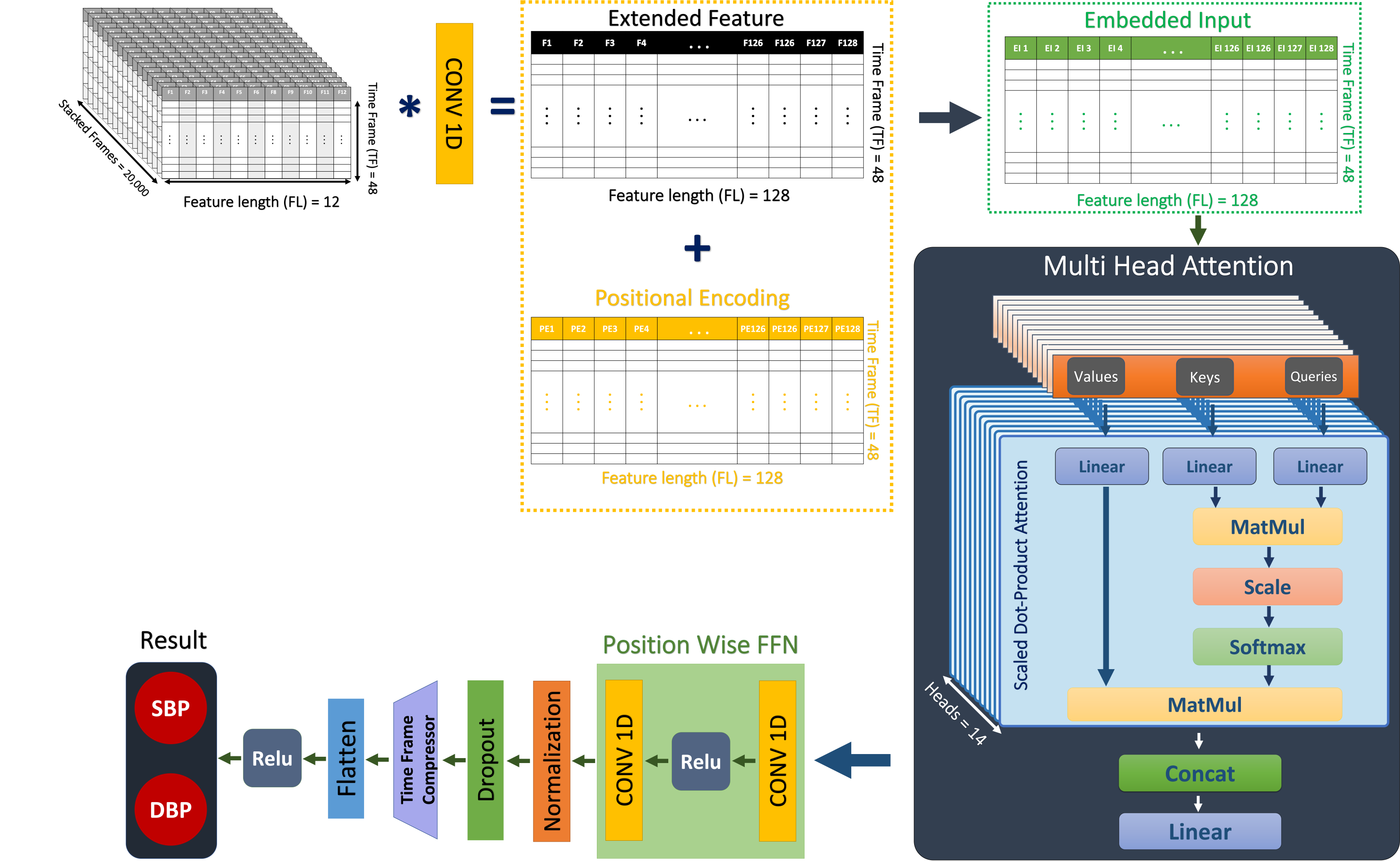}
\caption{\small This block diagram depicts the streamlined process for estimating systolic and diastolic blood pressures. Features are extracted from the dataset (top left), combined with positional encoding, and form an embedded input matrix. This matrix is then inputted into the model block, which integrates a multi-head attention mechanism. The output of the model is further refined in the position-wise block. Finally, the Time compressor and flattening blocks are applied to reduce the time frame and yield the estimated systolic and diastolic blood pressure values.}
\label{fig:blockD}
\end{figure}
This section focuses on material and methodology, whereas in the first part (the dataset formation subsection), we describe the preprocessing and feature extraction approaches. In the second part, the network architecture subsection provides an overview of the utilized modeling process. In summary and according to the workflow of our study, shown in Fig.~\ref{fig:blockD}, after dataset formation and preprocessing, the overall dataset is compiled based on $12$ final features. Next, the positional encoding values are added to the flattened values of the dataset, resulting in the construction of the embedded input matrix, which is used to feed the model. In the model part, a multi-head attention algorithm is employed. The position-wise process is applied to the output of the multi-head attention block, and finally, by reducing the time frame and flattening the position-wise output, the values related to systolic and diastolic blood pressures are obtained.

\subsection{Dataset Formation}
\begin{figure}[t!]
\centering
\includegraphics[width=0.7\linewidth]{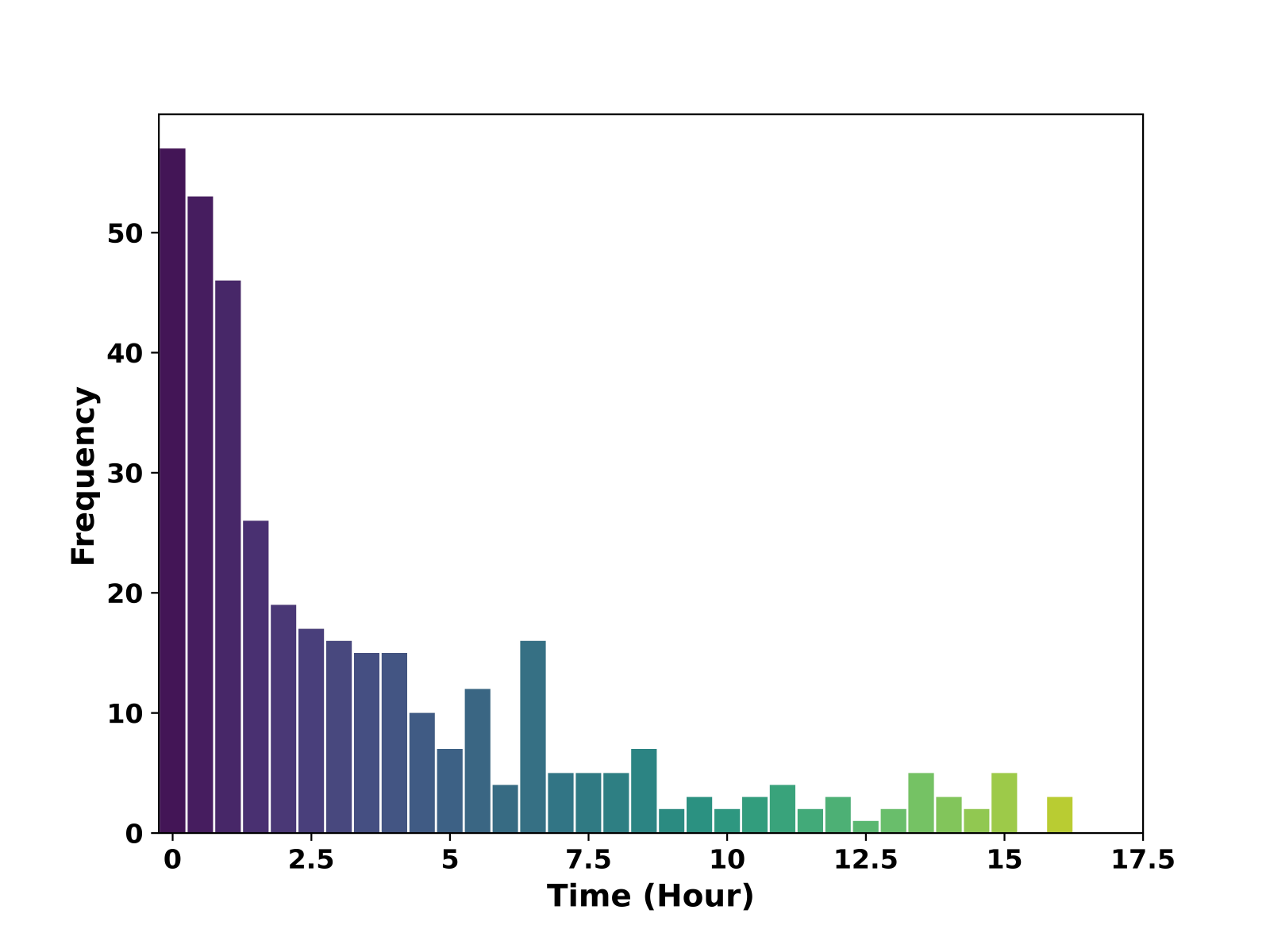}
\caption{\small Time distribution of records depicted in the graph. The $x-axis$ represents the record's length, ranging from a few minutes to a maximum of 17 hours. The darker shades of blue indicate a higher number of observations. The majority of patients have records with a time duration of less than 1.5 hours.}
\label{fig: t_dist}
\end{figure}

\begin{figure}[t!]
\centering
\includegraphics[width=1\linewidth]
{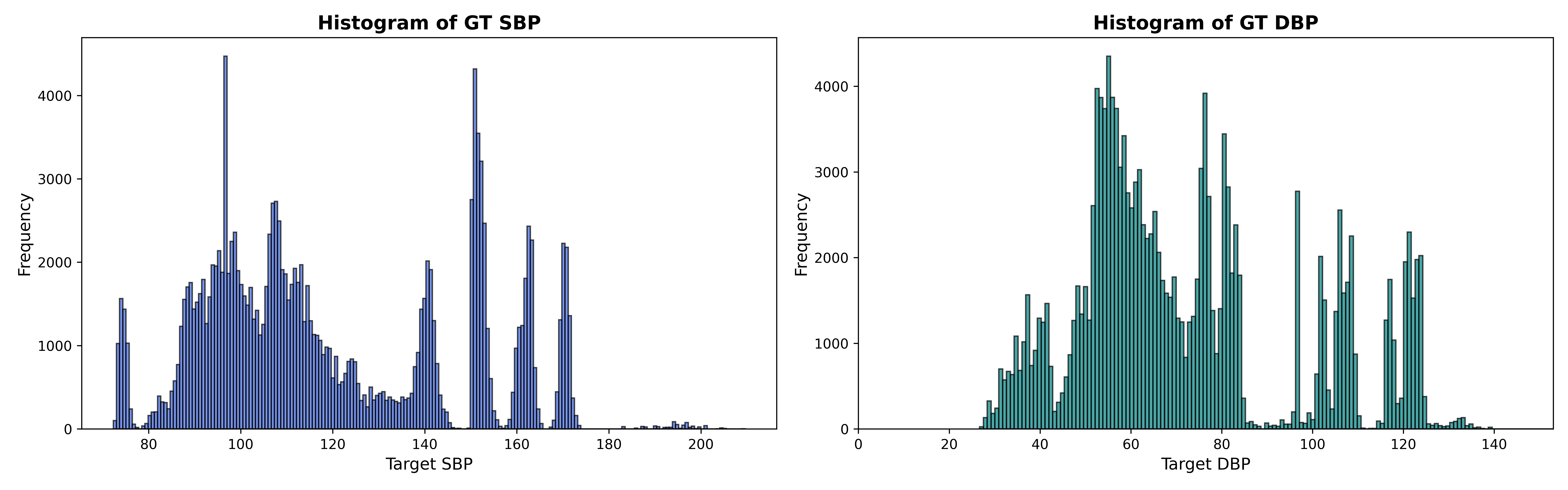}
\caption{\small Distribution of patient data based on SBP and DBP values. The $x-axis$ represents the blood pressure value, while the $y-axis$ depicts the data frequency for systolic blood pressure (left) and diastolic blood pressure (right). Each bar corresponds to the number of records of corresponding pressure. The density of the bars reveals the concentration of observations, highlighting that most patients fall within the normal range.}
\label{fig: sbp_dbp_dist}
\end{figure} 

The Medical Information Mart for Intensive Care (MIMIC)-IV Waveform Dataset is employed, which is available on the Physio-Net repository~\citep{38-mimiciv} to analyze the version of waveform recordings from bedside monitoring devices in modern Intensive Care Units (ICUs). MIMIC-IV includes ECG, PPG, and invasive ABP signals and is the latest and most recently released Dataset of its kind following a joint effort from Beth Israel Deaconess Medical Center (BIDMC) and Massachusetts Institute of Technology (MIT). The recording process was conducted at a sampling rate of $62.4$ Hz. All $198$ individuals in the dataset were completely de-identified to ensure patient confidentiality. Fig.~\ref{fig: t_dist} illustrates the duration of the recorded signals and their corresponding iterations. Most patients were monitored for less than $1.5$ hours, while some patients had multiple recordings lasting over $15$ hours.  
Although previous studies have employed MIMIC-II and MIMIC-III datasets for blood pressure estimation, to the best of our knowledge, this study stands out as the first utilization of the MIMIC-IV v2.0 dataset for this purpose. 
\subsubsection{Preprocessing}
\begin{figure}[t!]
\centering
\includegraphics[width=\linewidth]{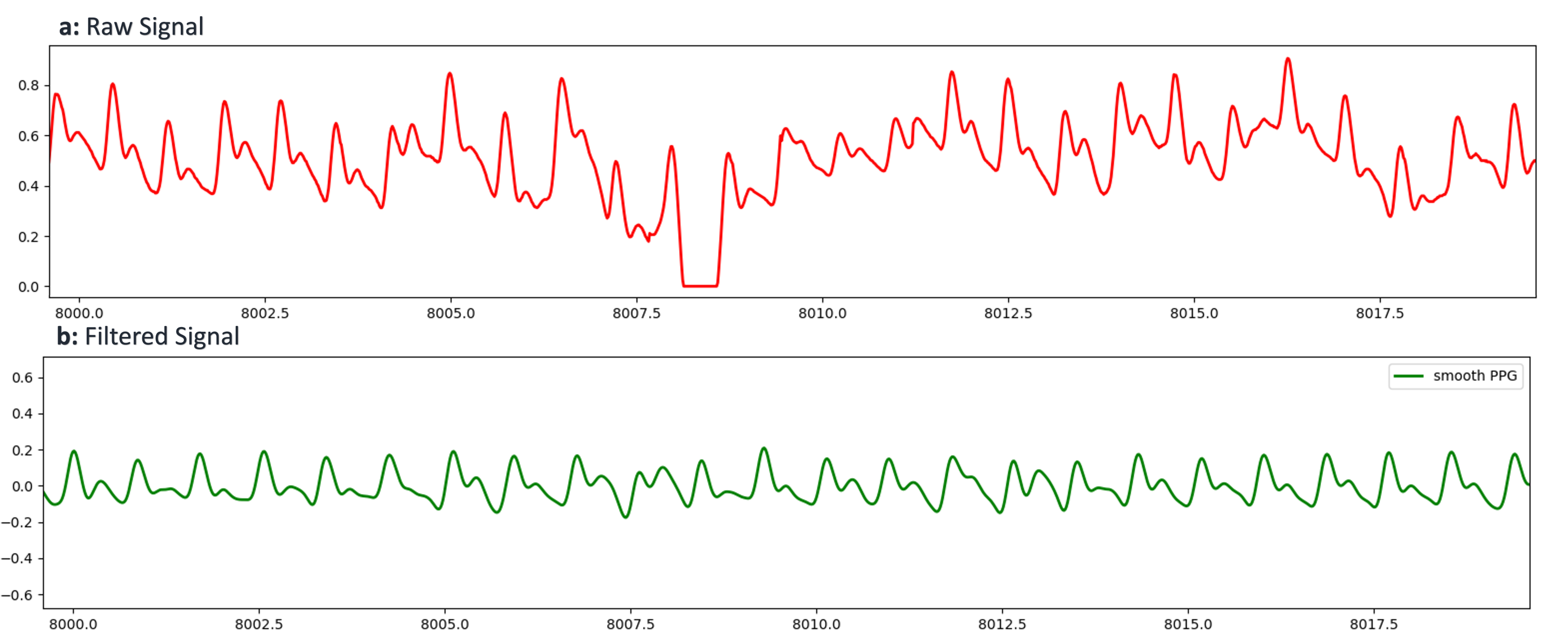}
\textcolor{orange}{\caption{\small A comparison of the signal before and after preprocessing. The x-axis represents time (s), while the y-axis denotes signal amplitude. The figure demonstrates the effectiveness of the preprocessing stage in suppressing noise and producing a cleaner and smoother signal, while accurately preserving the peaks and feet of the waveform.}}
\label{fig:signal_prep}
\end{figure}
\textcolor{orange}{ The preprocessing stage consisted of signal normalization, quality control, and filtering. Each retained PPG recording was independently normalized using recording-level z-score normalization. Records shorter than 15~min were excluded, and the normalized signals were divided into 10-s windows. A window was rejected when more than 10\% of its samples were outside the valid range of $-5$ to $+5$ standard-deviation units. Windows containing any 125-sample interval with a standard deviation below 0.02 or a peak-to-peak amplitude below 0.05 normalized units were also rejected as flat-line or insufficiently pulsatile. The complete preprocessing and signal-quality criteria are summarized in Table~\ref{tab:preprocessing_thresholds}. The MIMIC-IV Waveform Database used in this study contains 200 raw waveform records from 198 patients. After preprocessing and quality screening, the retained recordings were divided into 360 valid analysis segments. Each PPG signal was then processed using a fifth-order Butterworth bandpass filter with cutoff frequencies of 0.7 and 10~Hz. Zero-phase forward--backward filtering was applied to avoid phase displacement of waveform landmarks. A three-sample moving-average filter, corresponding to approximately 0.048~s at 62.4~Hz, was subsequently applied to reduce residual fluctuations. This procedure suppressed baseline drift and high-frequency noise while preserving the PPG morphology required for cycle detection and feature extraction. An example of the signal before and after preprocessing is presented in Figure~\ref{fig:signal_prep}. }

\subsubsection{Feature Extraction}

\begin{figure}[t!]
    \centering
    \includegraphics[width=\linewidth]{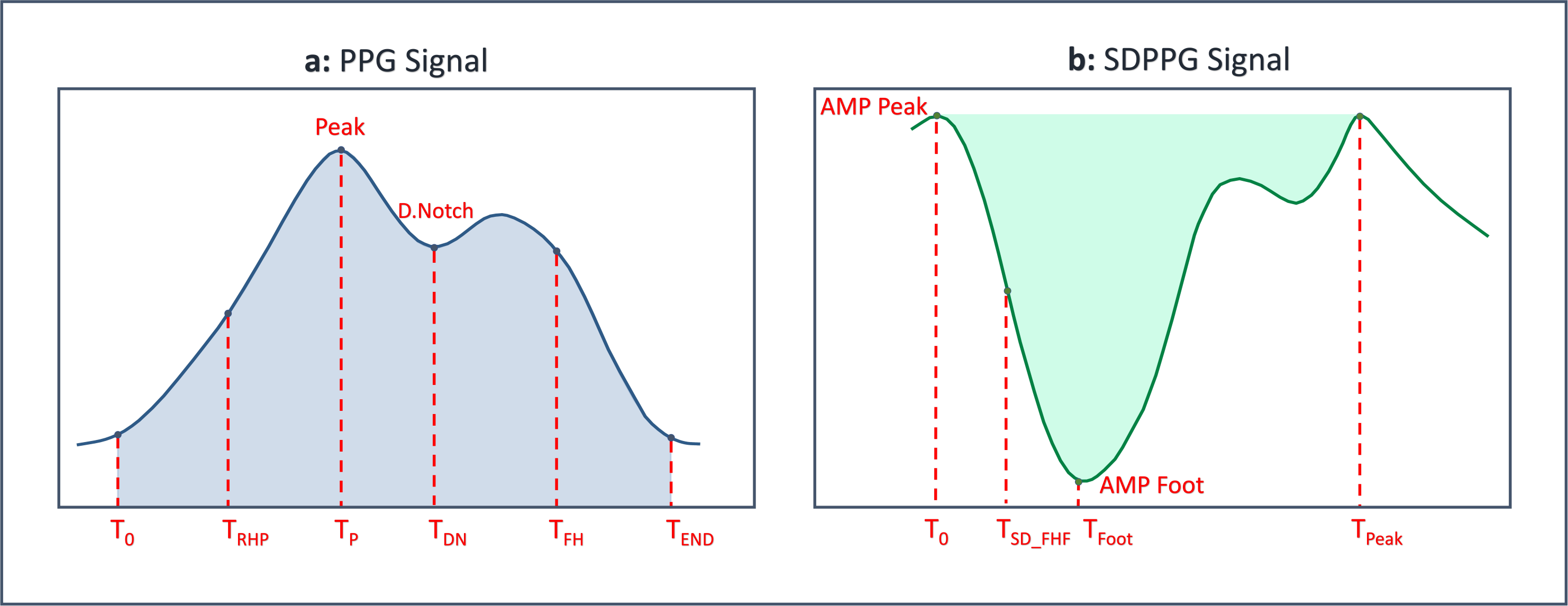}
    \caption{\small Characterization of features on the signal. In the PPG signal, the peak is defined as the maximum value of a PPG wave, while d.notch is defined as the minimum value between two consecutive peaks. In the SD PPG signal, amp peak refers to the maximum amplitude, while amp foot refers to the minimum amplitude between two consecutive waves.}
    \label{fig:features}
\end{figure}

\textcolor{orange}{ For blood-pressure estimation, the PPG waveform was analyzed at the cardiac-cycle level to capture morphological and temporal characteristics that may be obscured when complete 10-s windows are used directly~\citep{tazarv2021deep,slapnicar2019blood}. The feature-extraction procedure first identified the systolic peaks, cycle boundaries, and relevant fiducial points in the PPG waveform and its first (FDPPG) and second (SDPPG) derivatives. Signal-quality screening, filtering, peak-distance constraints, and cycle-duration criteria were applied according to the numerical parameters reported in Table~\ref{tab:preprocessing_thresholds}. A candidate pool of 24 cycle-level morphological and temporal features was then extracted from the PPG, FDPPG, and SDPPG signals. These features describe waveform amplitude, timing, slope, area, dicrotic-notch morphology, cycle duration, and derivative-based characteristics. The complete candidate feature set, denoted by F1--F24, is defined in Table~\ref{tab:features}, and representative waveform landmarks are illustrated in Figure~\ref{fig:features}. To prevent information leakage, feature selection was repeated independently within each patient-independent cross-validation fold using only the corresponding training patients. Pearson correlations between each candidate feature and the SBP and DBP targets were calculated using the training subset only~\citep{duan2016feature,el2020cuffless,putra2023study}. Features satisfying the significance criterion ($p<0.05$) were screened for redundancy, and highly collinear features were removed using a pairwise correlation threshold of $|r|>0.80$. Twelve features were retained in each fold and then applied, without reselection, to the corresponding held-out test patients. The resulting fold-specific feature sets are reported in Table~\ref{tab:patient_cv_features}.It is also worth mentioning that feature redundancy was addressed through fold-specific collinearity filtering. Additional experiments using all features and random 80\% subsets increased training cost or produced inconsistent results without meaningful performance improvement.  }

 \begin{table}[H]
	     \caption{\small Definition of the candidate features.}
	     \centering
	     \scriptsize
	     \renewcommand{\arraystretch}{1.15}
	     \setlength{\tabcolsep}{5pt}
	
	     \begin{tabularx}{\columnwidth}{llX}
		         \hline
		         \textbf{ID} & \textbf{Feature} & \textbf{Description} \\
		         \hline
		
		         F1  & PPG\_Integration
		         & Area under one PPG cycle. \\
		
		         F2  & PPG\_Pulse\_Amplitude
		         & Amplitude difference between the PPG peak and cycle start. \\
		
		         F3  & SDPPG\_Foot\_Peak
		         & Duration between the SDPPG foot and SDPPG peak. \\
		
		         F4  & PPG\_Rise\_Peak
		         & Duration from the start of the PPG cycle to the systolic peak. \\
		
		         F5  & FDPPG\_Positive\_Peak
		         & Maximum positive amplitude of the first-derivative PPG signal. \\
		
		         F6  & PPG\_Fall\_Half
		         & Duration from the PPG peak to the half-amplitude point in the falling phase. \\
		
		         F7  & SDPPG\_Extremum\_Amp
		         & Maximum amplitude of the SDPPG signal within one cycle. \\
		
		         F8  & PPG\_Notch\_Ratio
		         & Normalized amplitude ratio of the dicrotic notch. \\
		
		         F9  & PPG\_Cycle\_Duration
		         & Total duration of one PPG cycle. \\
		
		         F10 & PPG\_Peak\_To\_End
		         & Duration from the PPG peak to the end of the cycle. \\
		
		         F11 & FDPPG\_Negative\_Peak
		         & Minimum negative amplitude of the first-derivative PPG signal. \\
		
		         F12 & SDPPG\_Integration
		         & Area under the selected SDPPG curve region. \\
		
		         F13 & PPG\_Rise\_Half\_Time
		         & Duration from the start of the PPG cycle to the half-amplitude point in the rising phase. \\
		
		         F14 & PPG\_Rise\_Slope
		         & Average slope between the cycle start and the systolic peak. \\
		
		         F15 & PPG\_Cycle\_Count\_Frame
		         & Number of detected PPG cycles within one frame. \\
		
		         F16 & PPG\_Notch\_Depth
		         & Amplitude difference between the PPG peak and the dicrotic notch. \\
		
		         F17 & SDPPG\_Start\_To\_Fall\_Half
		         & Duration from the SDPPG start point to the half-amplitude point in the falling phase. \\
		
		         F18 & FDPPG\_Peak\_Time
		         & Duration from the PPG cycle start to the maximum positive FDPPG point. \\
		
		         F19 & PPG\_Peak\_To\_Notch
		         & Duration between the PPG peak and the dicrotic notch. \\
		
		         F20 & PPG\_Fall\_Slope
		         & Average slope of the falling phase after the systolic peak. \\
		
		         F21 & PPG\_Half\_Width
		         & Width of the PPG cycle at half of the peak amplitude. \\
		
		         F22 & PPG\_Rise\_Time\_Ratio
		         & Ratio between the rise time and total cycle duration. \\
		
		         F23 & FDPPG\_Peak\_To\_Trough
		         & Duration between the maximum positive and minimum negative FDPPG points. \\
		
		         F24 & PPG\_Notch\_Time\_Ratio
		         & Normalized temporal position of the dicrotic notch during the falling phase. \\
		
		         \hline
		     \end{tabularx}
	     \label{tab:features}
	 \end{table}


\begin{table}[t!]
	\caption{\footnotesize Patient-independent fold composition, observed
		hypertension and hypotension counts, and fold-specific feature selection
		after collinearity filtering.}
	\centering
	\tiny
	\renewcommand{\arraystretch}{1.05}
	\setlength{\tabcolsep}{2pt}
	
	\resizebox{\columnwidth}{!}{%
		\begin{tabular}{lcccccccp{3.2cm}}
			\hline
			\textbf{Fold} &
			\textbf{Train} &
			\textbf{Test} &
			\textbf{Overlap} &
			\textbf{Hyper.} &
			\textbf{Hypo.} &
			\textbf{SBP range} &
			\textbf{DBP range} &
			\textbf{Selected features after} \\
			&
			\textbf{patients} &
			\textbf{patients} &
			\textbf{patients} &
			\textbf{patients*} &
			\textbf{patients*} &
			\textbf{(mmHg)} &
			\textbf{(mmHg)} &
			\textbf{collinearity filtering} \\
			\hline
			
			Fold 1 & 158 & 40 & 0 & 6 & 3 & 72--216 & 32--138 &
			F1, F2, F3, F6, F7, F8, F11, F14, F15, F18, F20, F24 \\
			
			Fold 2 & 158 & 40 & 0 & 5 & 2 & 78--188 & 36--126 &
			F1, F4, F5, F7, F8, F12, F14, F15, F17, F18, F20, F24 \\
			
			Fold 3 & 158 & 40 & 0 & 7 & 4 & 70--224 & 34--136 &
			F2, F3, F5, F6, F8, F11, F12, F14, F15, F18, F20, F23 \\
			
			Fold 4 & 159 & 39 & 0 & 4 & 3 & 84--210 & 42--145 &
			F1, F3, F5, F7, F13, F15, F16, F18, F20, F21, F23, F24 \\
			
			Fold 5 & 159 & 39 & 0 & 6 & 4 & 76--196 & 30--124 &
			F2, F4, F6, F7, F8, F11, F12, F15, F18, F20, F23, F24 \\
			
			\hline
			\multicolumn{9}{p{0.95\textwidth}}{%
				\textit{*Hypertension was defined as SBP $\geq 140$~mmHg, and hypotension as SBP $< 90$~mmHg. F15, F18, and F20 were selected in all five patient-independent folds, indicating the highest selection stability across patient subsets.}
			}
		\end{tabular}%
	}
	
	\label{tab:patient_cv_features}
\end{table}


\begin{table*}[!t]
	\caption{\small Preprocessing, signal-quality, cycle-detection, and label-extraction thresholds.}
	\centering
	\scriptsize
	\renewcommand{\arraystretch}{1.25}
	\setlength{\tabcolsep}{6pt}
	\begin{tabular}{p{0.23\textwidth} p{0.72\textwidth}}
		\hline
		\textbf{Parameter} & \textbf{Value / criterion} \\
		\hline
		
		Minimum record duration &
		Records shorter than 15 min were excluded. \\
		
		PPG normalization and valid range &
		Each retained PPG recording was independently normalized before
		segmentation using
		\(z_i=(x_i-\mu_{\mathrm{record}})/\sigma_{\mathrm{record}}\),
		producing a signal with zero mean and unit standard deviation.
		The valid amplitude range was $-5$ to $+5$ standard-deviation units. \\
		
		Out-of-range rejection &
		After normalization and before filtering, a 10-s window was rejected
		when more than 10\% of its samples were outside the range $-5$ to
		$+5$, indicating excessive large-amplitude deviations or severe
		artifacts. \\
		
		Flat-line rejection &
		A window was rejected if any sliding interval of 125 samples
		(approximately 2.00 s at 62.4 Hz) had a standard deviation below
		0.02 or a peak-to-peak amplitude below 0.05 normalized units. \\
		
		Bandpass filtering &
		A fifth-order Butterworth bandpass filter with cutoff frequencies of
		0.7 and 10 Hz was applied using zero-phase forward--backward
		filtering. \\
		
		Moving-average smoothing &
		A three-sample moving average filter, corresponding to approximately
		0.048 s at 62.4 Hz, was applied after bandpass filtering. \\
		
		Peak and cycle detection &
		The minimum peak distance was 25 samples (approximately 0.40 s).
		Cycles with durations of 25--93 samples (approximately 0.40--1.49 s),
		corresponding to approximately 40--150 beats/min, were retained. \\
		
		ABP label extraction &
		SBP and DBP were defined as the maximum and minimum ABP values within
		each valid ABP cycle and averaged across valid cycles within the
		synchronized 10-s PPG--ABP window. Before averaging, cycles were
		required to satisfy SBP 70--250 mmHg, DBP 30--150 mmHg,
		SBP $>$ DBP, and pulse pressure 20--150 mmHg. \\
		
		\hline
	\end{tabular}
	\label{tab:preprocessing_thresholds}
\end{table*}


\subsubsection{Cycle-Level Frame Construction and Sequence Formation}

\textcolor{blue}{After preprocessing the PPG signals into 10-second windows, valid cardiac cycles were identified and converted into feature frames, each composed of 12 cycle-level features. The input construction pipeline therefore involved cycle detection, feature-frame generation, temporal ordering, and fixed-length sequence partitioning. After preprocessing, cycle extraction, and quality filtering, a total of \(N_i = 20{,}000\) valid cycle-level feature frames were obtained. These frames were subsequently grouped into fixed-length sequences for transformer-based learning. In the final design, the sequence length was set to \(T=48\), so that each transformer input contained 48 temporally ordered valid cycle-level frames. As shown in Table~\ref{tab:ablation_seq}, to support this design choice, an ablation study was conducted using \(T=24\), \(32\), \(48\), and \(64\). The results showed that \(T=48\) yielded the most favorable trade-off between predictive performance and computational efficiency, and was therefore selected for the final model.}
\begin{table}[t]
	\centering
	\caption{Ablation study on transformer input sequence length \(T\).}
	\label{tab:ablation_seq}
	\scriptsize
	\setlength{\tabcolsep}{3pt}
	\renewcommand{\arraystretch}{1.05}
	\begin{tabular}{lccccccccc}
		\hline
		\(T\) & MAE$_{\text{SBP}}$ & MAE$_{\text{DBP}}$ & RMSE$_{\text{SBP}}$ & RMSE$_{\text{DBP}}$ & GFLOPs & Lat. (ms) & Params (M) & Size (MB) \\
		\hline
		24 & 1.61 & 1.19 & 2.49 & 1.98 & 1.78 & 17.6 & 26.84 & 102.39 \\
		32 & 1.46 & 1.11 & 2.34 & 1.91 & 2.17 & 20.9 & 26.85 & 102.43 \\
		48 & \textbf{1.37} & \textbf{1.06} & \textbf{2.21} & \textbf{1.84} & \textbf{2.43} & \textbf{24.3} & \textbf{26.87} & \textbf{102.51} \\
		64 & 1.42 & 1.10 & 2.29 & 1.88 & 3.06 & 30.4 & 26.88 & 102.56 \\
		\hline
	\end{tabular}
\end{table}

\subsection{Network Architecture}
The proposed $\TR$ framework is developed based on the regressive time series transformer network, which is designed to learn from time series data while maintaining sequence order and enabling parallel computation during the training and inference phases. To enhance the dimensionality of the feature vectors, an embedding feature, previously implemented in Natural Language Processing (NLP) tasks~\citep{34-kalyan2020secnlp} is utilized. The initial structure of the features is defined by Stacked Frames (SF) based on shuffled frames, with $20,000$ frames ($N_i$) and a time frame sequence of $48$, which is given by
\begin{equation}
    SF=\left[F_T^{L, 1},F_T^{L, 2},\dots,F_T^{L, n}\right],
    \label{eq: SF}
\end{equation}
where $n$ refers to the number of frames, $F_T^{L, i}$ represents an individual frame that spans a shuffled frame with the input feature length of $L$ ($L_{in}=12$). Furthermore, $T$ refers to the time frame sequence and equals $48$. To extract the extended features, a $1$D convolution layer is applied to the stacked frames as follows
\begin{equation}
    ST_{Extended} (N_{i},L_{out} )=Bias(L_{out})+ \sum\limits_{k=0}^{L_{in}} weight(L_{out},k)input(N_i,k),
     \label{eq: ST}
\end{equation}
where $ST_{Extended}\in \R^{N_i\times L_{out}\times T}$, $weight\in \R^{L_{out}\times L_{in} \times KS}$, and  $input\in \R^{N_{i}\times L_{in} \times T}$. Equation.~\ref{eq: ST} expands the feature length from $12$ ($L_{in}$) to 128 ($L_{out}$) with a kernel size ($KS$) of 1.

\subsubsection{Multi-Head Attention Mechanism}
The self-attention mechanism captures the sequence dependencies and features carried on each frame. This mechanism employs three main matrices, i.e., Key, Queue, and Value, which are obtained by multiplying the input embedding matrix with learnable weights $W_K$, $W_Q$, and $W_V$, respectively. The resulting matrices are then used to compute a set of attention scores. To capture additional information about in-frame sequence dependency, multiple learnable sets (Heads) are stacked and concatenated, In the current study, we utilize a multi-head attention mechanism with $14$ heads for BP estimation. The architecture parameters of the networks such as heads-number are set regarding the experiments.
\textcolor{orange}{Additionally, the model incorporates a linear projection layer after the multi-head attention block to enhance feature transformation before the final prediction stage. This design allows the network to better model temporal and morphological dependencies among the cycle-level PPG features, including variations in pulse amplitude, waveform shape, and inter-cycle patterns, thereby improving blood pressure estimation performance.}

Despite the self-attention mechanism being a key component of the transformer architecture, it does not incorporate information about the relative positions of tokens in the input sequence. To address this limitation, a technique called Positional Encoding (PE) was introduced that encodes the position of each element in the input sequence by adding a sinusoidal function of different frequencies to the input embedding~\citep{17-wang2021cuff}. The PE array is added, in an element-wise fashion, to the $ST_{Extended}$ array, resulting in an Embedded input with a feature-length of $48$ and a time frame sequence length of $128$. The incorporated approach enables the transformer architecture to capture both the content and positional information of the input sequence.

{\color{orange}
	\vspace{.1in}
	\noindent
	\textbf{\textit{Position-Wise Feed-Forward Neural Network:}} The next component of the attention module is the position-wise feed-forward network, which is applied identically to each timestep in the input sequence. In the proposed transformer layer, this module consists of two linear projections: an expansion layer from \(256\) to \(8{,}192\), followed by a ReLU activation function, and a projection layer from \(8{,}192\) back to \(256\). This design follows the standard transformer architecture with an expansion factor of \(32\), enabling the model to capture complex nonlinear feature interactions within the temporal sequence. By transforming each position independently, the feed-forward network enhances the learned representation of each cycle-level feature vector and helps the model better encode context-relevant information. In addition, normalization and dropout layers are incorporated to improve generalization and reduce the risk of overfitting~\citep{35-vaswani2017attention}.}

{\color{orange}
	\vspace{.1in}
	\noindent
	\textbf{\textit{Time Frame Compressor and Flattening:}} 
	In the initial stages of experimentation, computational challenges were encountered due to the high overhead associated with processing long temporal representations on local hardware. In particular, the final flattening stage, which converts the high-dimensional sequence output into a one-dimensional representation for regression, was identified as a computational bottleneck. To address this issue, a time-frame compressor was introduced before flattening in order to reduce the temporal dimensionality while preserving the most informative sequential patterns. Specifically, the time-frame compressor is implemented as a dense interpolation layer with a fixed quadratic weighting kernel. Its input is the encoder output sequence of length \(T=48\), where each timestep is represented by a feature vector of dimension \(d_{\text{model}}=256\). The module maps this input sequence into a compressed temporal representation with 32 output positions using a pre-computed weight matrix \(W\), which is not updated through gradient descent. In this way, the temporal dimension is reduced from 48 to 32, while the feature dimension remains unchanged. The entries of \(W\) are constructed using quadratic distance-based functions, so that each output position is formed as a smooth weighted combination of neighboring input timesteps. The output of this module is therefore a compressed sequence of size \(32 \times 256\), which preserves informative temporal patterns for the subsequent regression stage. Thereafter, in the final stage of the model, the compressed sequence is flattened and passed to the regression head to estimate systolic and diastolic blood pressure values.}
%

\subsection{Cross-Validation}

\textcolor{orange}{ To evaluate the proposed $\TR$ framework, patient-independent 5-fold cross-validation was performed using patient IDs as the splitting unit. All records, segments, and frames belonging to the same patient were retained within a single fold, ensuring zero patient overlap between the training and test sets. In each iteration, approximately 80\% of the patients were used for training and the remaining 20\% were held out for testing. Feature selection was repeated independently within each fold using only the training patients. Pearson correlation screening ($p<0.05$) was followed by removal of highly collinear features ($|r|>0.80$). Data scaling, hyperparameter selection, and model fitting were also restricted to the training data, and the held-out patients and their labels were not accessed during model development. The patient composition, blood-pressure ranges, and fold-specific selected features are reported in Table~\ref{tab:patient_cv_features}. Performance was calculated independently on each held-out fold and summarized as the mean and standard deviation across the five folds. }

\begin{table*}[t!]
	\caption{\small Patient-independent 5-fold cross-validation performance of TransfoRhythm.}
	\label{tab:patient_cv_performance}
	\centering
	
	\tiny
	\renewcommand{\arraystretch}{1.05}
	\setlength{\tabcolsep}{1.5pt}
	
	\resizebox{0.92\textwidth}{!}{%
		\begin{tabular}{lccrrrrrrrr}
			\hline
			
			\textbf{Fold} &
			\shortstack{\textbf{Test}\\\textbf{patients}} &
			\shortstack{\textbf{Test}\\\textbf{samples}} &
			\multicolumn{4}{c}{\textbf{SBP}} &
			\multicolumn{4}{c}{\textbf{DBP}} \\
			
			\cline{4-7}
			\cline{8-11}
			
			& & &
			\textbf{MAE} &
			\textbf{RMSE} &
			\textbf{ME} &
			\textbf{R-square} &
			\textbf{MAE} &
			\textbf{RMSE} &
			\textbf{ME} &
			\textbf{R-square} \\
			
			\hline
			
			Fold 1 & 40 & 3,850 &
			1.73 & 3.44 & $+0.11$ & 0.874 &
			1.64 & 2.55 & $-0.21$ & 0.881 \\
			
			Fold 2 & 40 & 4,120 &
			1.90 & 2.58 & $+0.86$ & 0.918 &
			1.40 & 2.44 & $-0.38$ & 0.887 \\
			
			Fold 3 & 40 & 3,940 &
			2.08 & 3.32 & $-0.77$ & 0.871 &
			1.30 & 2.80 & $-0.36$ & 0.872 \\
			
			Fold 4 & 39 & 4,050 &
			1.86 & 3.16 & $-0.38$ & 0.890 &
			1.40 & 2.71 & $-0.23$ & 0.879 \\
			
			Fold 5 & 39 & 4,040 &
			2.08 & 3.07 & $+0.14$ & 0.887 &
			1.72 & 2.46 & $-0.12$ & 0.936 \\
			
			\hline
			
			\textbf{Mean} &
			\textbf{198} &
			\textbf{20,000} &
			\textbf{1.93} &
			\textbf{3.11} &
			\textbf{$-0.01$} &
			\textbf{0.888} &
			\textbf{1.49} &
			\textbf{2.59} &
			\textbf{$-0.26$} &
			\textbf{0.891} \\
			
			\hline
		\end{tabular}%
	}
\end{table*}


\subsection{Evaluation Metrics}
To evaluate the effectiveness of the proposed $\TR$ framework, we have identified four key evaluation metrics, i.e., R-squared~($ R^{2}$), Mean  Error ($ME$), Mean Absolute Error ($MAE$), and Root Mean Square Error ($RMSE$), given by
\begin{align}
R^{2} &= \frac {\sum\limits_{i=1}^{n}\left(T_{i}-P\right)^2} {\sum\limits_{i=1}^{m}\left(T_{i}-\Bar{T}\right)^2 }\label{eq: R},\\
ME &= \frac{1}{n}\sum\limits_{i=1}^{n}(T_{i}-P_{i}) \label{eq: ME},\\
MAE &= \frac{1}{n}\sum\limits_{i=1}^{n}|T_{i}-P_{i}| \label{eq: MAE},\\
RMSE &= \sqrt{\frac{1}{n}\sum\limits_{i=1}^{n}(T_{i}-P_{i})^2 }. \label{eq: RMSE}
\end{align}
where $T_i$ refers to the target and $P_{i}$ is the predicted value. In brief, Metric $ R^{2}$ serves as an indicator of the extent to which the variance in the dependent variable can be attributed to the independent variables. The other metrics ($ME$, $MAE$, and $RMSE$)  quantify the magnitude and/or percentage of errors between actual and predicted values. 

\section{Results}\label{sec:results}
\begin{table}[t!]
    \caption{Values of the considered hyperparameters }
    \centering
    \begin{tabular}{ll}
        \hline
        \textbf{Hyperparameters}	&	\textbf{Type and Value}	\\ \hline
        Dropout	&	0.15	\\
        Optimization Algorithm	&	Adam \\
        Learning Rate	&	1E-4 (with dynamic wight decay)	\\
        Activation Function	&	ReLU \\
        Loss Function	&	MSE	\\
        Batch Size	&	128	\\
        Epoch	&	400 \\ \hline
    \end{tabular}
    \label{tab:hyperp}
\end{table}
To evaluate the proposed $\TR$ framework, comprehensive experiments were conducted by varying hyperparameters and modifying architectural components. The finalized set of hyperparameters is outlined in Table~\ref{tab:hyperp}. Training was performed for 400 epochs on a local server equipped with an Intel $12700$ CPU, 16 GB of RAM, and an Nvidia RTX 3060 GPU with 12 GB of VRAM. Performance metrics, including $R^2$, mean error (ME), mean absolute error (MAE), and root mean square error (RMSE), were calculated based on cross-validation results as detailed in the methodology section.

\textcolor{orange}{ Under the patient-independent 5-fold cross-validation protocol, $\TR$ achieved mean $R^2$ values of $0.888$ for SBP and $0.891$ for DBP. The corresponding MAE values were $1.93$~mmHg and $1.49$~mmHg, while the RMSE values were $3.11$~mmHg and $2.59$~mmHg for SBP and DBP, respectively. These results demonstrate consistent predictive performance across the five held-out patient groups. Detailed fold-level results are provided in Table~\ref{tab:patient_cv_performance}. }

\subsection{Error Investigation}
\begin{figure}[t!]
    \centering
    \includegraphics[width=\linewidth]{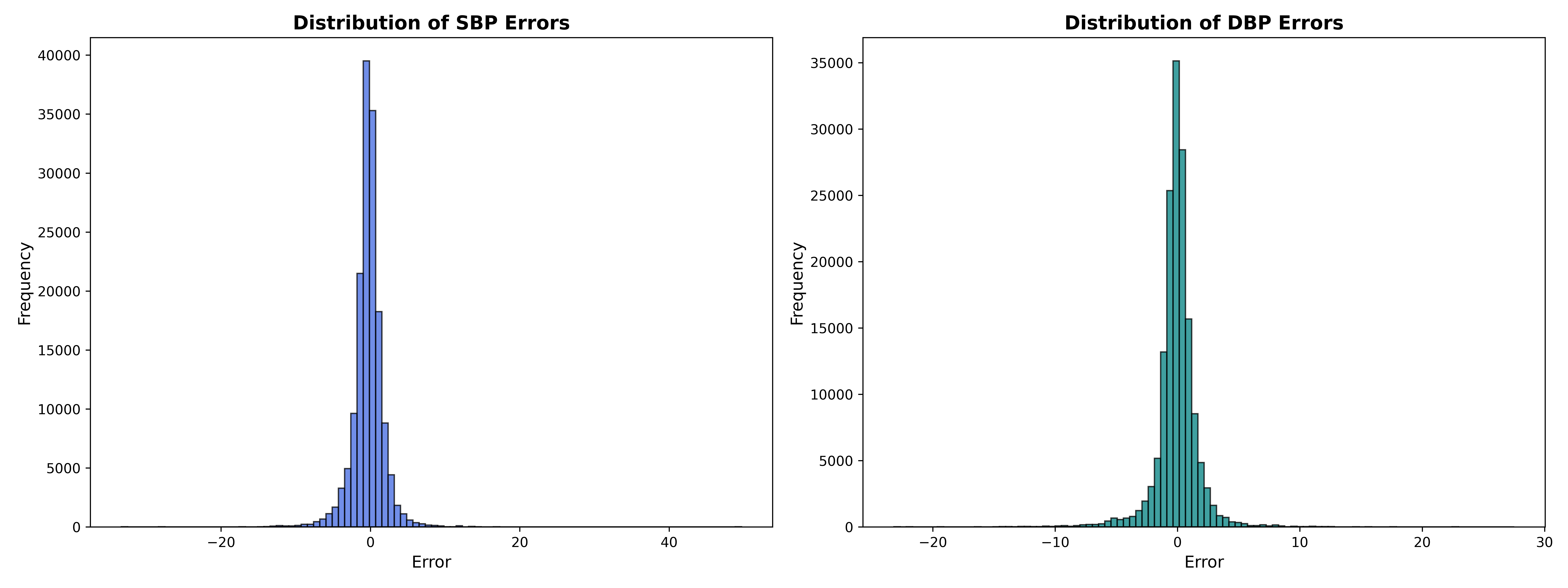}
    \caption{\small Distributions of prediction errors for systolic (SBP) and diastolic (DBP) blood pressure measurements. The error histograms are centered around zero, indicating balanced model performance for SBP and DBP.}
    \label{fig:distribution_errors}
\end{figure}
\begin{figure}[t!]
    \centering
    \includegraphics[width=\linewidth]{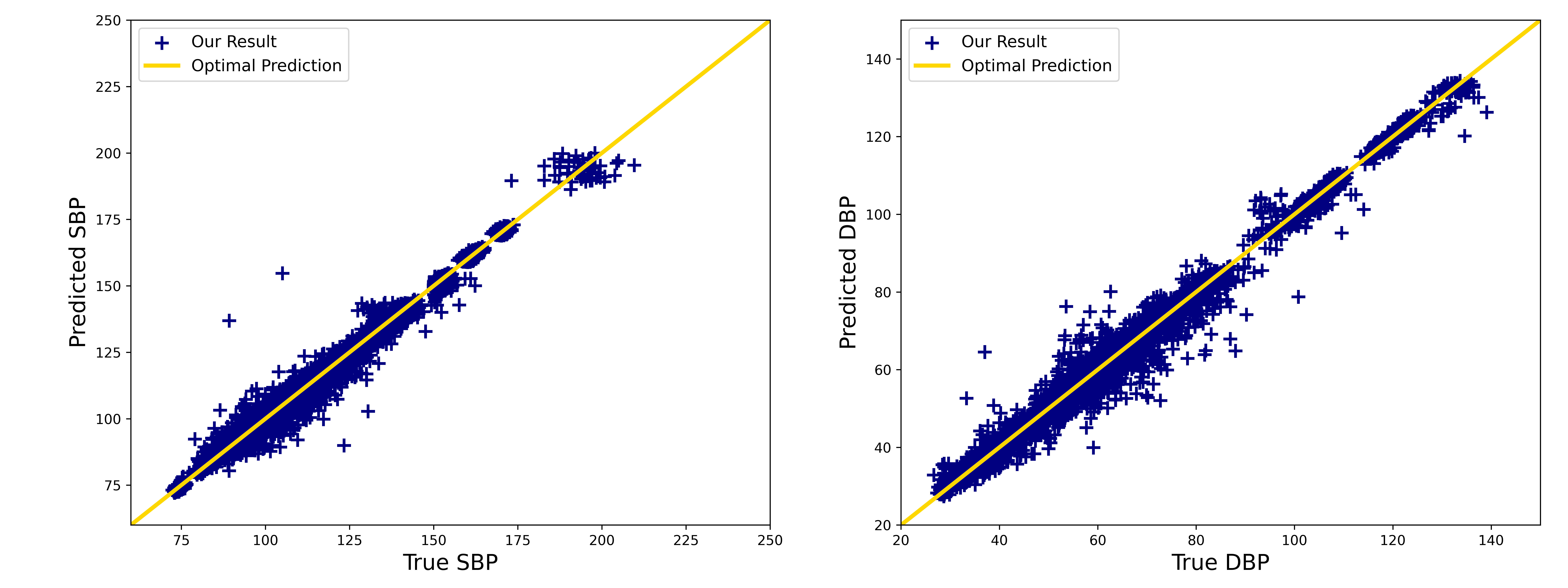}
    \caption{\small True versus predicted blood pressure values for systolic blood pressure (left) and diastolic blood pressure (right). The data points (navy color) indicate model predictions, while the golden line represents the ideal correspondence between true and predicted values.}
    \label{fig:predicted_versus_true_values}
\end{figure}
\begin{figure}[t!]
    \centering
    \includegraphics[width=\linewidth]{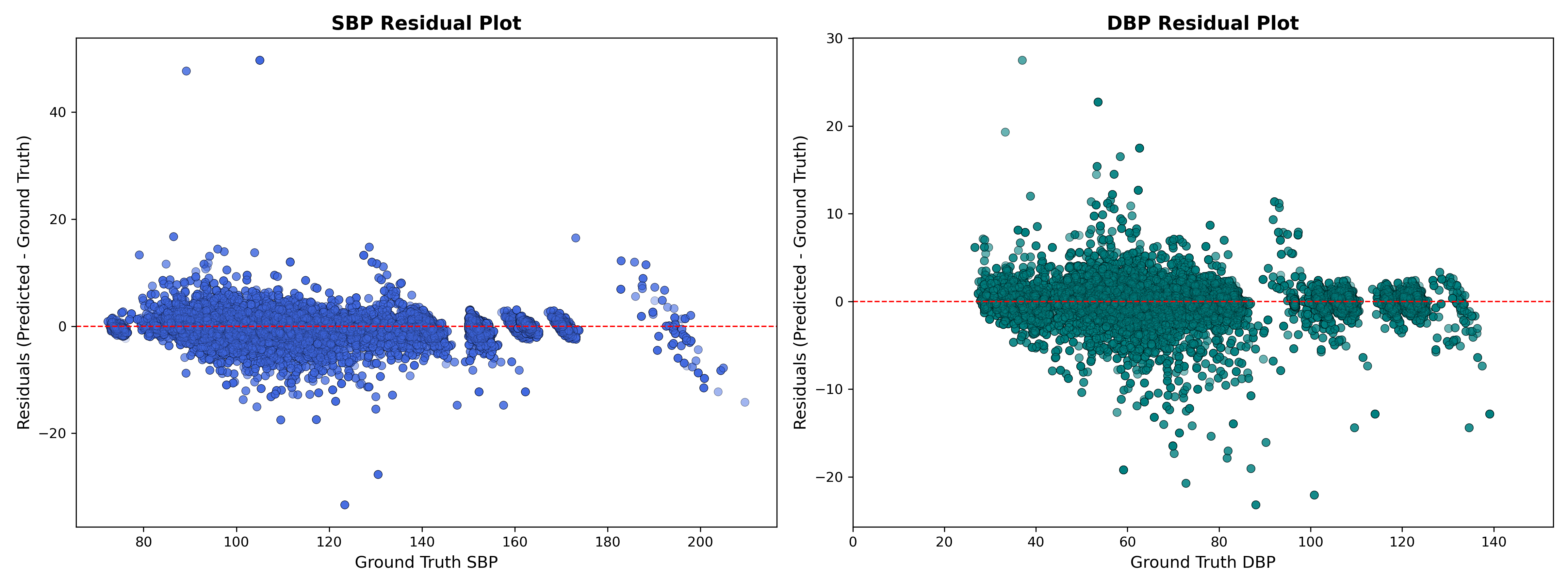}
    \caption{\small Residual plots for systolic (SBP) and diastolic (DBP) blood pressure predictions. The dashed red line represents zero residuals, emphasizing a balanced error distribution around this reference.}
    \label{residual}
\end{figure}

Understanding the behavior of prediction errors and their alignment with ground truth values is essential for evaluating the reliability and robustness of the proposed $\TR$ framework. Figure~\ref{fig:distribution_errors} presents the distribution of errors for both systolic (SBP) and diastolic (DBP) blood pressure predictions. The histograms are centered around zero, indicating that the model's performance is well-balanced, with no significant bias toward over- or underestimation. The relatively narrow spread of errors suggests that most predictions are in close agreement with the true values, highlighting the precision of the model. In Figure~\ref{fig:predicted_versus_true_values}, the relationship between true and predicted values is shown, with the left plot representing SBP and the right plot showing DBP. Each data point reflects a model prediction, and the golden line represents the ideal one-to-one correspondence. {\color{orange}
	The alignment of the predictions with the identity line indicates good
	agreement across the blood-pressure range represented in the evaluated
	MIMIC-IV samples. Generalizability beyond this dataset remains to be established
	through external validation.
}. Furthermore, the residual plots in Figure~\ref{residual} provide additional insights by illustrating the differences between predicted and true values. The residuals are evenly distributed around a dashed red line representing zero error, indicating a lack of systematic bias. {\color{orange}
The residual distribution indicates limited systematic bias and consistent
performance across the evaluated samples.
} {\color{orange}
Together, these analyses indicate low prediction error and limited average
bias under the current evaluation protocol.
}

\subsection{Comparison with AAMI and BHS Numerical Accuracy Criteria}

\begin{table}[t!]
	\caption{Evaluation of the Proposed Model with AAMI and BHS Standards.}
	\centering
	\scriptsize 
	\renewcommand{\arraystretch}{1.1}
	
	\begin{tabular}{m{2.5cm}m{1.5cm}m{1cm}m{1cm}m{1cm}m{1cm}m{1cm}m{1cm}}
		\hline
		
		& & 
		\multirow{2}{*}{\textbf{Records}} &
		\multirow{2}{*}{\textbf{\parbox{1cm}{ME\\(mmHg)}}} &
		\multirow{2}{*}{\textbf{\parbox{1cm}{STD\\(mmHg)}}} &
		\multicolumn{3}{c}{\textbf{Cumulative error percentage}} \\
		
		\cline{6-8}
		
		& & & & &
		\textbf{<5 mmHg} &
		\textbf{<10 mmHg} &
		\textbf{<15 mmHg} \\
		
		\hline
		
		\textbf{$\TR$ Architecture} &
		SBP & 360 & -0.01 & 3.1 &
		93.28\% & 96.83\% & 98.63\% \\
		
		& DBP & 360 & -0.26 & 2.57 &
		94.20\% & 96.14\% & 97.18\% \\
		
		\hline
		
		\textbf{AAMI Standard Requirement} &
		& >85 & <5 & <8 & - & - & - \\
		
		\hline
		
		\multirow{3}{2.5cm}{\textbf{BHS Standard Requirement}} &
		Grade A & - & - & - & 60\% & 85\% & 95\% \\
		
		& Grade B & - & - & - & 50\% & 75\% & 90\% \\
		
		& Grade C & - & - & - & 40\% & 65\% & 85\% \\
		
		\hline
	\end{tabular}
	
	\label{tab:comparison}
\end{table}

To ensure that the proposed $\TR$ architecture meets the international standards for medical device regulation, we evaluated our findings using the standards of the Association for the Advancement of Medical Instrumentation (AAMI) and The British Hypertension Society (BHS). To comply with AAMI standards, certain statistical conditions must be met, including a sample size of at least $85$ individuals aged $12$ years or older, a Mean Error (ME) of less than $5$ mmHg, and a Standard Deviation (SD) of the error less than $8$ mmHg. Additionally, BHS grades accuracy based on the cumulative percentage of error. Table~\ref{tab:comparison} provides a comparison between our results and these standard metrics.   As presented in Table~\ref{tab:comparison} the proposed model in this study has successfully met the AAMI standard. Moreover, the model has also achieved ``Grade A'' in the BHS standard. These results confirm the reliability and outstanding performance of the proposed model based on the prediction error values.

\subsection{Statistical Assessment via Bland-Altman}
\begin{figure}[t!]
    \centering
    \includegraphics[width=\linewidth]{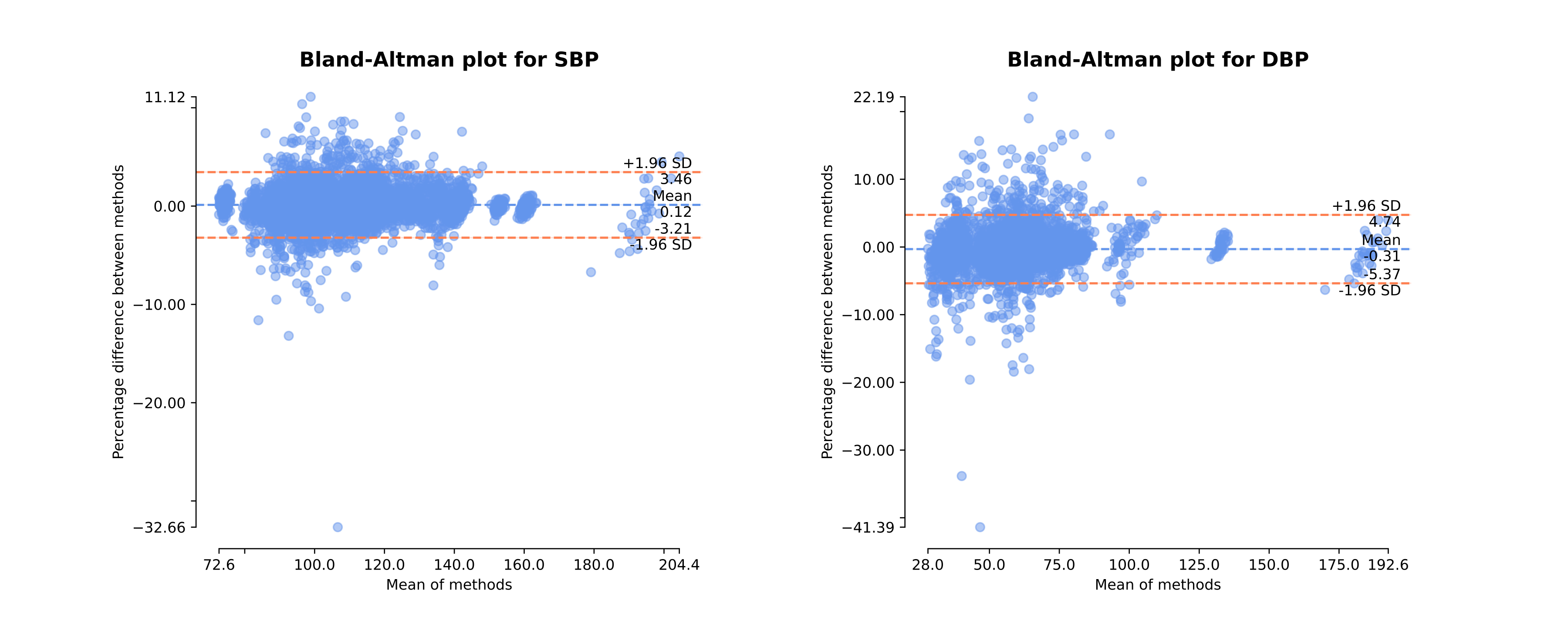}
    \caption{\small Bland-Altman plots for assessing agreement between ground truth and predicted values of SBP and DBP.}
    \label{fig: BAP}
\end{figure}
{\color{orange} The Bland-Altman method is a statistical approach to evaluate the agreement between two measurement techniques by plotting their differences against their mean. This method provides a visual and quantitative assessment of bias and the range within which the two methods agree. In Figure.~\ref{fig: BAP}, the Bland-Altman plot for the $\TR$ model demonstrates a mean difference of $0.26$ mmHg for SBP, with $95\%$ limits of agreement ranging from $-3.60$ mmHg to $4.12$ mmHg. For DBP, the mean difference is $0.07$ mmHg, with $95\%$ limits of agreement spanning $-5.81$ mmHg to $5.96$ mmHg. These results indicate that the $\TR$ model has minimal systematic bias and maintains a relatively narrow range of agreement. Additionally, the random distribution of differences around the mean suggests the absence of proportional bias, highlighting the model's reliability. 
	This analysis indicates good agreement between the model predictions and the
	invasive ABP reference values within the evaluated dataset. Further external
	and prospective studies are required before considering clinical application.
}

\subsection{Comparisons with Benchmark Models}
This section focuses on validating our results by comparing the performance of our proposed model with benchmark algorithms from previous studies. Given the lack of prior work utilizing the MIMIC IV dataset for BP prediction, we implemented benchmark models from other studies that analyzed earlier versions of the MIMIC dataset. These benchmark models are employed to establish a baseline performance for blood pressure estimation in the context of the MIMIC IV dataset. The selection of benchmark models was based on specific criteria. Firstly, we considered models that exclusively utilized PPG signals. Furthermore, due to the impracticality of conducting extensive feature engineering for each model, we specifically chose models that were trained on raw data. Additionally, we selected models with open-access codes with satisfying results in their respective papers. The benchmark models enclosed various architecture families. From CNN-based networks, we selected ResNet1D and AlexNet1D, which were originally implemented and evaluated on MIMIC II and MIMIC-III datasets~\citep{schrumpf2021assessment}. Additionally, we included a U-Net model that had been evaluated on the MIMIC III dataset~\citep{ibtehaz2022ppg2abp}. 
In terms of Recurrent Neural Networks (RNN), we defined a BiLSTM model for our benchmark set, which aimed to address the absence of suitable RNN-based models meeting our selection criteria. The BiLSTM model begins with a 1D Convolutional layer for feature extraction, followed by a Max pooling layer to enhance pattern detection. Four Bidirectional LSTM layers then capture contextual understanding by considering past and future states. To prevent overfitting, four Dropout layers are placed after the LSTM layers. Finally, two dense layers are used to perform the final regression task and map features to the desired output format. By containing the BiLSTM model, we aimed to gain insights into RNN-based approaches for blood pressure prediction and expand the scope of evaluation by comparing our model performance with other benchmark models. Lastly, we incorporated a hybrid architecture combining CNN and RNN layers, which utilized the raw PPG signal along with its first and second derivatives of the MIMIC III dataset as input features~\citep{slapnivcar2019blood}.
It is essential to note that the pre-processing methods employed in the benchmark studies were customized for their specific MIMIC dataset version. As the structure of the MIMIC IV dataset differs from the earlier versions, we utilized our pre-processing pipelines. Additionally, since the referenced studies employed a $10-second$ window size, we maintained consistency by adopting the same window length for the data.

\textcolor{orange}{Table~\ref{tab:comparison_bench_final} } presents the quantitative evaluation metrics obtained by implementing the benchmark models on the MIMIC IV dataset. Each benchmark model was trained using different sets of hyperparameters, and the most accurate model was identified and reported.
{\color{orange}
	Under the same patient-independent evaluation protocol, TransfoRhythm
	achieved lower errors than the reimplemented benchmark models for most reported
	metrics.
}

\textcolor{blue}{ For a fair comparison, all benchmark models in Table~\ref{tab:comparison_bench_final} were re-implemented and evaluated using the same MIMIC-IV data, patient-level folds, and evaluation protocol as $\TR$. Statistical comparisons were repeated using patient-independent out-of-fold prediction errors. Paired $t$-tests and Wilcoxon signed-rank tests were used to assess the observed differences ~\citep{bland2010statistical,cuzick1985wilcoxon}. $\TR$ achieved statistically significant improvements over several baseline models and metrics ($p<0.005$ for the corresponding comparisons), although significance was not observed in every case. In particular, the DBP MAE difference between $\TR$ and ResNet\_1D was not statistically significant. These results support the consistency of the observed improvements while limiting superiority claims to the comparisons supported by the statistical tests. }


\begin{table}[t!]
	\caption{Comparison with benchmark models on the MIMIC-IV dataset.}
	\label{tab:comparison_bench_final}
	\centering
	\scriptsize
	\setlength{\tabcolsep}{3pt}
	\renewcommand{\arraystretch}{1.15}
	\resizebox{\linewidth}{!}{%
		\begin{tabular}{llcccccccccc}
			\hline
			\textbf{Arch.} & \textbf{Model} & \textbf{MAE} & \textbf{MAE} &
			\textbf{RMSE} & \textbf{RMSE} & \textbf{$R^2$} & \textbf{$R^2$} &
			\textbf{Params} & \textbf{GFLOPs} & \textbf{Latency} & \textbf{Size} \\
			
			& & \textbf{SBP} & \textbf{DBP} &
			\textbf{SBP} & \textbf{DBP} &
			\textbf{SBP} & \textbf{DBP} &
			\textbf{(M)} & & \textbf{(ms)} & \textbf{(MB)} \\
			
			\hline
			
			CNN & ResNet\_1D &
			2.28 & 1.06 &
			5.55 & 3.21 &
			0.962 & 0.985 &
			3.85 & 0.86 & 11.00 & 14.68 \\
			
			CNN & AlexNet\_1D &
			3.09 & 1.59 &
			7.98 & 4.60 &
			0.922 & 0.969 &
			2.35 & 0.13 & 7.30 & 8.99 \\
			
			CNN & U-Net &
			2.76 & 1.56 &
			7.47 & 4.70 &
			0.931 & 0.967 &
			68.07 & 1.532 & 48.50 & 259.66 \\
			
			RNN & Bi-LSTM &
			3.86 & 1.90 &
			7.98 & 4.19 &
			0.922 & 0.974 &
			2.56 & 0.231 & 17.80 & 9.76 \\
			
			Hybrid & CNN-RNN Network &
			2.79 & 1.96 &
			5.62 & 3.91 &
			0.961 & 0.977 &
			13.28 & 0.319 & 19.36 & 50.65 \\
			
			\hline
			
			\textbf{MHA} & \textbf{TransfoRhythm} &
			\textbf{1.93} & \textbf{1.49} &
			\textbf{3.11} & \textbf{2.59} &
			\textbf{0.888} & \textbf{0.891} &
			\textbf{26.87} & \textbf{2.418} &
			\textbf{24.15} & \textbf{102.51} \\
			
			\hline
		\end{tabular}%
	}

	\vspace{0.3em}
	\raggedright\scriptsize
	All complexity metrics were obtained from the authors' re-implementations under the same profiling setup, using the same input configuration as in the experiments, FP32 precision, batch size of 1, and single-core core-i9 CPU  inference.
\end{table}
\section{Discussion and Conclusion}\label{sec:conclu}
Long-term BP measurement could not be carried out using cuff-based solutions because of its indiscrete nature and the inaccuracy caused by factors such as the cuff size, placement, and the patient's movement. Consequently, there has been a surge of recent interest in using DL techniques to estimate BP based on ECG and/or PPG data~\citep{36-malayeri2022concatenated}. In this context,  the $\TR$ framework is developed in this study for continuous BP estimation. The $\TR$ takes as input $12$ features extracted from the MIMIC IV raw and derivatives PPG waveforms. It is worth noting that despite many studies on the MIMIC datasets, to the best of our knowledge, this work is the first investigation of BP prediction implemented on the MIMIC IV waveform version dataset.  
At the core of the $\TR$ framework is a Multi-Head Attention (MHA)-based mechanism, which is a neural network architecture with several advantages over its traditional DL methods for time series analysis~\citep{42-wu2019pay}. Several studies have shown that MHA can outperform conventional DL methods in various time series prediction tasks. For example, MHA-based networks have been shown to outperform RNNs in predicting fault diagnosis~\citep{48-jiang2022bearing}, energy consumption assessment~\citep{bu2020time}, and speech enhancement~\citep{nicolson2020masked}.
Additionally, MHA has been shown to improve prediction accuracy in stock price prediction~\citep{50-chen2020multi}, and traffic flow prediction~\citep{51-reza2022multi}. When it comes to BP estimation, conventional DL methods are typically developed based on RNN and/or CNN architectures. While these methods have shown promising results, they have limitations, such as difficulty in capturing long-term dependencies or invariance to temporal shifts in the input~\citep{43-blohm2018comparing} ~\citep{nicolson2020masked}. The primary advantage of MHA is its ability to attend to various input sequence segments, enabling the model to learn complicated relationships between various observations in the sequence and improve representations of the input. It, therefore, can capture both short-term and long-term dependencies in time series data, where detailed relations between distinct observations might be difficult to capture by conventional DL methods~\citep{44-li2022attention} ~\citep{45-niu2024attention}. Attention-based networks have also the potential to enhance forecasting accuracy and improve the robustness of time series analysis by handling noise and perturbations~\citep {45-niu2024attention}. Another merit of the MHA is its ability to handle variable-length inputs without using reduction methods, making it possible to learn more accurate representations of the input. In conventional DL methods, padding or truncation is often used to ensure all input sequences have the same length. This could lead to information loss and reduced model performance~\citep{bu2020time}. Finally, the Attention mechanism can also be used to improve the interpretability of the model in diverse academic contexts \citep{li2021attention}. Attention scores can provide insight into the specific segments of the input sequence that the model focuses on when making predictions ~\citep{47-ahmed2022transformers}. This can be extremely practical in BP studies, in which identifying important observations can provide valuable insight. 


\hyphenpenalty=10000
\begin{table}[t!]
\centering
\scriptsize
\renewcommand{\arraystretch}{1.4}
\caption{Comparison of studies in BP estimation in terms of methodology, database, features, and metrics (NR means not reported).}
\resizebox{\textwidth}{!}{%
\begin{tabular}{
    >{\centering\arraybackslash}m{2.5cm}
    >{\centering\arraybackslash}m{1.0cm}
    >{\centering\arraybackslash}m{1.5cm}
    >{\centering\arraybackslash}m{2.0cm}
    >{\centering\arraybackslash}m{1.0cm}
    >{\centering\arraybackslash}m{1.0cm}
    >{\centering\arraybackslash}m{1.0cm}
    >{\centering\arraybackslash}m{1.0cm}
    >{\centering\arraybackslash}m{1.0cm}
    >{\centering\arraybackslash}m{1.0cm}
    >{\centering\arraybackslash}m{1.0cm}
    >{\centering\arraybackslash}m{1.0cm}}
\hline
\textbf{Model Architecture} & \textbf{Signal} & \textbf{Datasets} & \textbf{Input Type} & \textbf{MAE SBP} & \textbf{MAE DBP} & \textbf{RMSE SBP} & \textbf{RMSE DBP} & \textbf{STD SBP} & \textbf{STD DBP} & \textbf{ME SBP} & \textbf{ME DBP} \\
\hline
MLPlstm-BP~\citep{14-huang2022mlp} & ECG PPG & MIMIC~II  & Raw Waveform & 3.52 & 2.13 & 5.1 & 3.13 & 5.09 & 3.07 &  0.071 & -0.495 \\ \hline
Temporal Convolution Network (TCN)~\citep{1-zabihi2022bp} & ECG PPG &  MIMIC~I MIMIC~III & Feature Extraction & 2.59 & 1.33 & 3.03 & 1.58 & 3.58 & 1.97 & -2.98 & -3.65 \\ \hline
CNN + LSTM~\citep{13-baker2021hybrid}& ECG PPG & MIMIC~III  & Raw Waveform & 4.41 & 2.91 & NR & NR & 6.11 & 4.23 & NR & NR \\ \hline
NABNet (BiConvLSTM)~\citep{23-mahmud2023nabnet}& ECG PPG & MIMIC~III  & Raw Waveform (multi-channel) & 2.63 & 1.09 & NR & NR & 2.96 & 1.05  & -0.678 &  -0.342\\ \hline
Bi-LSTM~\citep{chen2023beat} & ECG PPG & MIMIC II & Feature Extraction & 2.51 & 1.38 & 3.22 & 1.78 & NR & NR & -0.09 & -0.07  \\ \hline
LSTM network~\citep{guo2024cuffless} & ECG PPG& MIMIC~III & Raw Waveform & 8.34 & 3.47 & NA & NA & 10.68 & 4.76 & NA & NA \\ \hline	

BiSpectral MRRNet \citep{subramanian2026bispectral} & ECG PPG & Primary Dataset + MIMIC~III & Spectro-temporal Representation & 3.780 & 2.428 & NR & NR & 6.868 & 4.619 & -1.002 & -0.670 \\ \hline

Deep Autoencoder~\citep{22-qin2021deep}& PPG & MIMIC~II  & Raw Waveform & 5.42 & 3.14 & NR & NR & 6.64 & 3.74 & 1.65 & -1.28\\ \hline
1D\_CNN + 2D\_CNN~\citep{36-malayeri2022concatenated}& PPG & MIMIC~II & Raw Waveform + Image & 3.05 & 1.58 & NR & NR & 5.26 & 2.6 &  -0.15 & -0.29 \\ \hline
CNN (Transfer Learning)~\citep{17-wang2021cuff}& PPG & MIMIC~II & Image (Transformed PPG) & 6.17 & 3.66 & 8.46 & 5.36 & 8.46 & 5.36 & 0.00 & -0.04 \\ \hline
Bi-GRU +GRU+attention ~\citep{32-el2021deep}& PPG & MIMIC~II  & Feature Extraction (22 features) & 2.58 & 1.26 & NR & NR & 3.35 & 1.63 & -0.52 & -0.66 \\ \hline
BiLSTM + LSTM + Attention~\citep{27-el2021cuffless}& PPG & MIMIC~II  & Feature Extraction & 4.51 & 2.6 & NR & NR & 7.81 & 4.41 & -0.48 &  -0.49\\ \hline
GRU~\citep{25-el2020cuffless} & PPG & MIMIC~II & Feature Extraction (7~features) & 3.25 & 1.43 & NR & NR & 4.76 & 1.77 & NR & NR \\ \hline
DNN~\citep{26-hsu2020generalized}& PPG & MIMIC~II & Feature Extraction (32~features) & 3.21 & 2.23 & 4.64 & 3.3 & NR & NR & NR & NR \\
\hline
U-Net~\citep{ibtehaz2022ppg2abp} & PPG & MIMIC~III & Raw Waveform & 5.73 & 3.45 & NR & NR & 9.16 & 6.14 & -1.582 & 1.619 \\ \hline
Feed Forward ANN~\citep{samimi2023ppg} & PPG & MIMIC~II UCI  & Feature Extraction (21 features) & 7.41 & 3.32 & NR & NR & 10.40 & 4.89 & -4.02 & -0.31 \\ \hline
CNN+LSTM~\citep{mahardika2023ppg} & PPG & MIMIC~III  & Feature Extraction & 3.79 & 2.89 & NR & NR & 7.89 & 5.34 & 0.13 & 0.48 \\ \hline
ArterialNet Transformer ~\citep{huang2023arterialnet}& PPG &  MIMIC~III & Raw waveform & 4.15 & 3.17 & 5.26 & 4.01 & 1.32 & 1.37 & NR & NR \\ \hline
KD-Informer~\citep{ma2022kd}& PPG & MIMIC~II & Backward elimination feature selection & 4.00 & 2.97 & NR & NR & 5.93 & 3.87 & 0.018 & 0.010 \\ \hline 
MLM-Transformer w/personalization ~\citep{chu2023non}& PPG &  MIMIC~III & Raw waveform & 2.41 &  1.31 &  NR & NR & 2.72 & 1.77 & 0.037 & 0.029\\ \hline
Transformer network~\citep{nawaz2024cufflessarterialbloodpressure}& PPG &  UCI  & Raw waveform & 3.77 & 2.69 & NR & NR & NR & NR & NR & NR \\ \hline
Self-supervised transformer~(STP)~\citep{ma2024stp}& PPG &  MIMIC~III & Raw waveform & 3.37 & 2.48 & NR & NR & 4.21 & 2.76 &  0.85 & 0.49 \\ \hline

U-Net + multiRes\citep{zhao2024amrunet} & PPG & UCI & Raw Waveform & 2.85 & 1.79 & NR & NR & NR & NR & -0.89 & 1.06 \\ \hline

CNN + Transformer\citep{tian2025paralleled} & PPG & MIMIC-III & Raw Waveform & 4.44 & 2.36 & NR & NR & 5.98 & 3.22 & 0.17 & 0.18 \\ \hline

TCN + Attention\citep{dai2024noninvasive} & PPG & UCI & Raw Waveform (Derivatives) & 5.3 & 2.11 & NR & NR & 8.3 & 3.17 & NR & NR \\ \hline

ResNet + BiLSTM\citep{liu2025two} & PPG & MIMIC~IV & Feature Extraction (60 features) & 3.47 & 2.81 & NR & NR & 5.06 & 4.11 & NR & NR \\ \hline

ACFA (xLSTM + Transformer)\citep{li2026acfa} & PPG & MIMIC~III & Raw Waveform  & 2.21 & 3.48 & NR & NR & 3.48 & 4.24 & NR & NR \\ \hline

%

\textbf{$\TR$} & \textbf{PPG} & \textbf{MIMIC~IV} & \textbf{Feature Extraction} & \textbf{1.93} & \textbf{1.49} & \textbf{3.11} & \textbf{2.59} & \textbf{3.1} & \textbf{2.57} & \textbf{-0.01} & \textbf{-0.26} \\ \hline
\label{tab:all-methods}
\end{tabular}%
}
\end{table}
\hyphenpenalty=50

\begin{figure}[t]
	\centering
	\includegraphics[width=\linewidth]{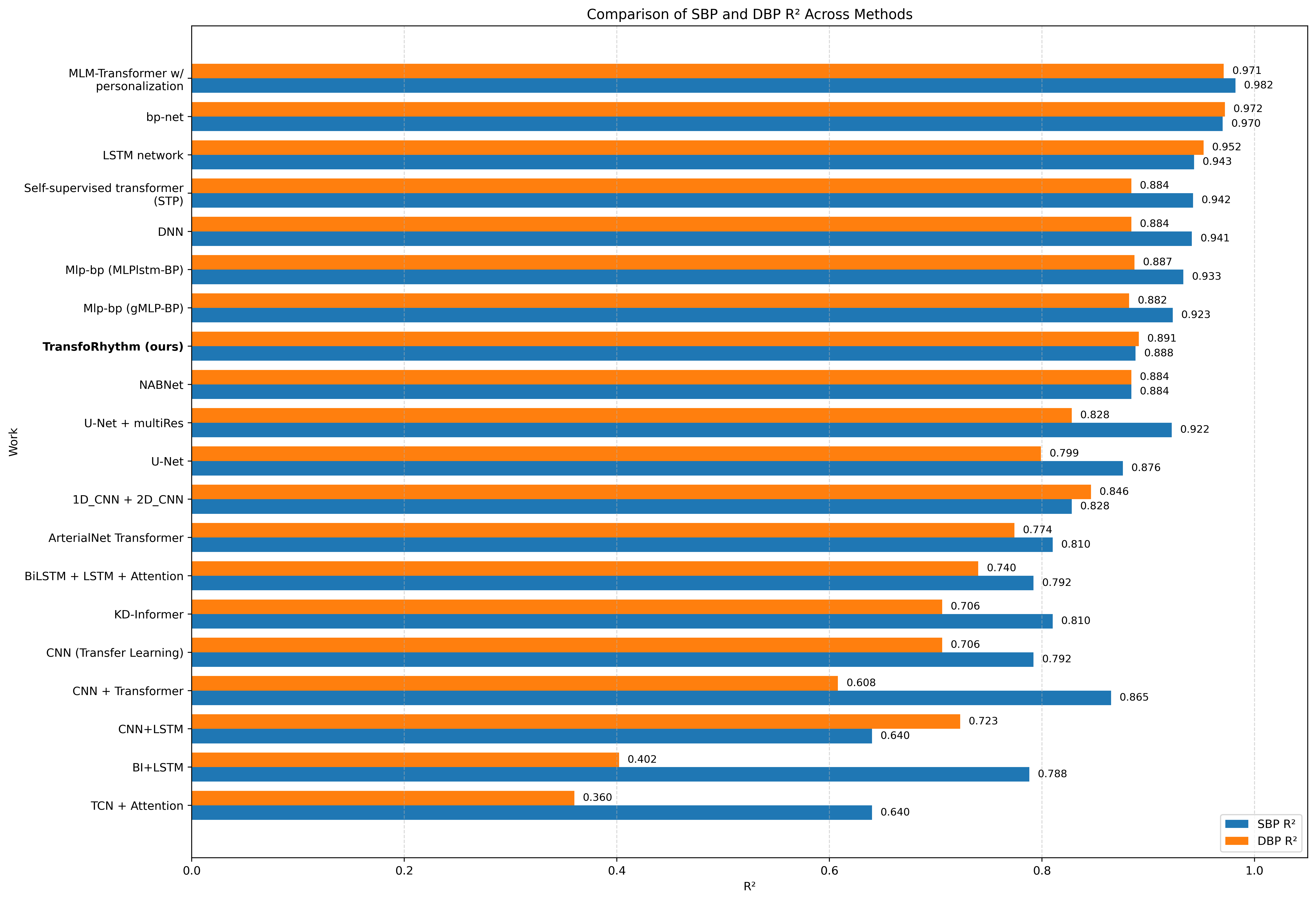}
	\caption{Comparison of SBP and DBP $R^2$ across methods.}
	\label{fig:r2-comparison}
\end{figure}

\textcolor{blue}{Figure~\ref{fig:r2-comparison} compares the SBP and DBP $R^2$ values obtained using the evaluated methods. The results indicate that \TR\ achieves $R^2$ values comparable to those of the benchmark models, demonstrating competitive predictive performance under the present experimental conditions.}

\textcolor{orange}{
	Our results, summarized in Table~\ref{tab:patient_cv_performance}, demonstrate the strong and consistent performance of the proposed $\TR$ framework under patient-independent 5-fold cross-validation. Across the five held-out patient folds, the model achieved MAE values ranging from $1.73$ to $2.08$~mmHg for SBP and from $1.30$ to $1.72$~mmHg for DBP. The corresponding mean MAE values were $1.93$~mmHg and $1.49$~mmHg, with mean RMSE values of $3.11$~mmHg and $2.59$~mmHg for SBP and DBP, respectively. Mean $R^2$ values of $0.888$ for SBP and $0.891$ for DBP further indicate strong agreement between the estimated and reference BP values. In addition, the mean errors of $-0.01$~mmHg for SBP and $-0.26$~mmHg for DBP show minimal systematic bias. Importantly, samples were not excluded solely on the basis of normotensive, hypotensive, or hypertensive BP categories. Only cycles outside broad physiological plausibility limits were rejected, thereby retaining clinically relevant low- and high-BP cases and avoiding an artificially restricted target distribution. The limited variation in performance across the five patient-disjoint folds therefore provides evidence of robust generalization within the evaluated MIMIC-IV cohort. To further examine the effectiveness of the proposed framework, two complementary evaluation stages were conducted.
}

\begin{itemize} \item \textbf{\textit{Benchmarks on the MIMIC IV Dataset:}} \ In the first stage, the proposed model's performance on the MIMIC IV dataset was compared against benchmark models trained on the same dataset, as summarized in Table~\ref{tab:comparison_bench_final}. This comparative analysis ensures a fair evaluation and reinforces confidence in the model's accuracy and reliability. {\color{orange}
		 TransfoRhythm achieved lower errors than the evaluated baselines. Its
		multi-head attention mechanism is designed to model temporal relationships
		among PPG cycles; however, the contribution of individual architectural
		components should be interpreted based on dedicated ablation experiments.
	} 
	By selectively attending to critical features, the MHA mechanism enables $\TR$ to effectively capture temporal dependencies and spatial relationships within PPG signals, enhancing its ability to focus on informative patterns for precise BP estimation.
\textcolor{orange}{ The benchmark results reveal differences among the evaluated architectural families. ResNet1D achieved the second-best overall performance, suggesting that residual one-dimensional convolutions effectively capture local PPG waveform patterns while supporting stable optimization. The hybrid CNN--RNN model~\citep{slapnivcar2019blood} and U-Net~\citep{ibtehaz2022ppg2abp} also produced competitive results by combining local feature extraction with temporal or multi-scale representations, although their additional architectural complexity did not yield lower errors than ResNet1D. AlexNet1D showed that a simpler convolutional architecture can provide reasonable performance, whereas Bi-LSTM, despite modeling bidirectional temporal dependencies, produced higher errors than the convolutional and hybrid baselines. Overall, the comparison suggests that local morphological representation is important for PPG-based BP estimation, while $\TR$ achieved the lowest overall errors by combining selected cycle-level morphological features with cross-cycle attention. However, this interpretation should be made cautiously because the models use architecture-specific input representations. }
%
\item \textbf{\textit{Comparisons with the State-of-The-Art Models:}} 
In the second stage of performance evaluation, the model's functionality is investigated by comparing it to related state-of-the-art studies. Table~\ref{tab:all-methods} illustrates the comparisons, which include model architecture, input signal type, dataset, feature injection model, and metrics. \textcolor{blue}{ Also, Figure~\ref{fig:r2-comparison} shows the comparison of SBP and DBP $R^2$ values across methods, reflecting their predictive performance}. The main objective is to provide a comprehensive evaluation of the model's performance, independent of dataset consideration. This analysis allows for a thorough assessment of the model's functionality and its relative standing compared to other advanced approaches in the field of BP estimation. Regarding datasets, most of the similar works employed MIMIC II or MIMIC III datasets. MIMIC-IV, which was used in this study in comparison with other available MIMIC datasets, includes more PPG data. Meanwhile, MIMIC-IV includes more recent data, which may be more representative of current clinical practices, and has undergone rigorous quality control processes to ensure that the data is accurate and reliable~\citep{38-mimiciv}. 
\textcolor{orange}{ Existing PPG-based BP-estimation methods can be grouped into three broad modeling directions. Traditional ML approaches, including Support Vector Machines (SVM)\citep{39-zhang2017svm}, Random Forest (RF)\citep{40-he2016beat}, and XGBoost~\citep{41-che2019continuous}, rely on engineered features and offer comparatively simple predictive pipelines, although their performance depends strongly on the quality of the selected physiological descriptors. Deep-learning methods reduce this dependence by learning representations from time-series signals. CNN-based models primarily capture local waveform morphology~\citep{17-wang2021cuff}, whereas LSTM- and GRU-based models focus on temporal dependencies across successive observations~\citep{25-el2020cuffless}. Hybrid architectures combine these complementary strengths by jointly modeling morphological and temporal information~\citep{32-el2021deep,27-el2021cuffless,30-rastegar2023hybrid}; for example, Malayeri \textit{et al.}\citep{36-malayeri2022concatenated} used a CNN-based design to capture both local and recursive signal patterns, while other studies integrated recurrent layers with attention mechanisms to emphasize informative temporal representations\citep{32-el2021deep,27-el2021cuffless}. Overall, the literature shows a progression from feature-dependent conventional models toward architectures that jointly capture waveform morphology and temporal context. However, differences in input representation, datasets, and validation protocols limit direct comparison across studies; therefore, Table~\ref{tab:all-methods} is presented as a contextual overview rather than a strict head-to-head evaluation. }
\textcolor{orange}{
	Recent Transformer-based blood pressure estimation methods follow several complementary directions. ArterialNet~\citep{huang2023arterialnet} and MLM-Transformer~\citep{chu2023non} learn temporal representations directly from raw PPG waveforms, whereas the Self-Supervised Transformer (STP)~\citep{ma2024stp} uses pretraining to improve representation learning. Hybrid approaches, such as the CNN--Transformer architecture in~\citep{tian2025paralleled}, combine convolutional extraction of local waveform morphology with attention-based modeling of longer-range temporal relationships. Collectively, these studies suggest that Transformer-based models can capture BP-related information through raw-waveform learning, self-supervision, or hybrid feature extraction. However, differences in datasets, input representations, personalization strategies, and validation protocols limit direct comparison and leave the generalization of these methods to unseen patients insufficiently established.
}

\textcolor{blue}{
	Recent advances in cuffless BP estimation increasingly exploit complementary signal representations. Liu et al.~\citep{liu2025two} fused SVM-RFE-selected PPG features with raw-waveform learning and reported SBP/DBP MAEs of 3.47/2.81~mmHg on MIMIC-IV. ACFA~\citep{li2026acfa} combined DyCASNet, Transformer, xLSTM, and FKAN modules and achieved subject-wise SBP/DBP MAEs of 2.21/3.48~mmHg on MIMIC-III. However, integrating multiple advanced deep-learning modules increases the model size and computational complexity. BiSpectral MRRNet~\citep{subramanian2026bispectral} used ECG--PPG STFT spectrograms with stacked ensemble learning and obtained subject-wise SBP/DBP MAEs of 3.780/2.428~mmHg. These studies demonstrate the potential of multimodal and multi-domain feature fusion, although direct comparisons remain limited by differences in datasets, model complexity, and validation protocols.
}
\end{itemize}

\textcolor{blue}{
\begin{itemize}
	\item \textbf{\textit{Hyperparameter tuning:}} Hyperparameters of the proposed Transformer model were optimized using Optuna\citep{akiba2019optuna} with its TPE-based Bayesian search strategy, with the objective of minimizing the validation MSE for ABP prediction. The search included the number of attention heads (\texttt{n\_heads}), compression factor (\texttt{factor}), number of transformer layers (\texttt{num\_layers}), learning rate (\texttt{lr}), and optimizer type. The best Optuna trial was then used for final training and evaluation.
\end{itemize}}

Despite the strong performance of the proposed $\TR$ framework, several limitations and future research opportunities exist. The MIMIC-IV dataset was used exclusively for training and validation, and all findings are based on this dataset. Thus, this study represents the first attempt to develop a BP estimation model solely based on PPG signals from MIMIC-IV. Future work could investigate the model's performance on additional datasets or incorporate diverse data sources to improve generalization. Finally, exploring more advanced attention-based network architectures could further enhance model accuracy and robustness.

\vspace{-.1in}
\section*{Acknowledgments}
This Project was partially supported by the Natural Sciences and Engineering Research Council (NSERC) of Canada through the NSERC Discovery Grant RGPIN-2023-05654 and RGPIN-262177-12.
\vspace{-.1in}
\section*{Data Availability}
The utilized dataset (Reference~\citep{38-mimiciv}) is publicly available through the following link: \href{www.physionet.org/content/mimic4wdb/0.1.0/}.
\vspace{-.1in}

\section*{Author Contributions Statement}
\textbf{Amir Arjomand and Amin Boudesh}: Conceptualization, Methodology, Software, Investigation, Data curation, Validation, Visualization, Writing – original draft. 

\textbf{Farnoush Bayatmakou}: Formal analysis, Resources, Writing – review and editing.  

\textbf{Georgiy Krylov}:Supervision,review and editing. 

\textbf{Kenneth B. Kent}: Supervision, Project administration, Writing – review and editing. 

\textbf{Arash Mohammadi}: Supervision, Project administration, Conceptualization, Methodology, Writing – review and editing.


\section*{Additional Information}
\textbf{Competing Interests}: Authors declare no competing interests.

\bibliographystyle{elsarticle-num}
\bibliography{References}       
\end{document}